\documentclass[%
 aip,
 amsmath,amssymb,
 reprint,%
]{revtex4-1}

\usepackage{graphicx}
\usepackage{dcolumn}
\usepackage{bm}

\usepackage[normalem]{ulem}
\usepackage[utf8]{inputenc}
\usepackage[T1]{fontenc}
\usepackage{mathptmx}

\usepackage{amsmath}
\usepackage{amsfonts}
\usepackage{graphicx}
\usepackage{amssymb}
\usepackage{bbold}
\usepackage{scalerel}
\usepackage{xcolor}

\newcommand{\dz}{\text{d}z}
\newcommand{\dr}{\text{d}\textbf{r}}
\newcommand{\dx}{\text{d}x}
\newcommand{\dy}{\text{d}y}

\newcommand{\ra}{\textbf{r}}

\newcommand{\comment}[1]{}

\begin{document}

\preprint{AIP/123-QED}

\title[]{Machine-learning free-energy functionals using density profiles from simulations}

\author{Peter Cats}
 \affiliation{Institute for Theoretical Physics,  Utrecht University, Princetonplein 5, 3584 CC Utrecht, the Netherlands}

\author{Sander Kuipers}
 \affiliation{Institute for Theoretical Physics,  Utrecht University, Princetonplein 5, 3584 CC Utrecht, the Netherlands}

\author{Sacha de Wind}
 \affiliation{Institute for Theoretical Physics,  Utrecht University, Princetonplein 5, 3584 CC Utrecht, the Netherlands}

\author{Robin van Damme}%
\affiliation{Soft Condensed Matter, Debye Institute for Nanomaterials Science, Princetonplein 1, 3584 CC Utrecht, the Netherlands}

\author{Gabriele M. Coli}
\affiliation{Soft Condensed Matter, Debye Institute for Nanomaterials Science, Princetonplein 1, 3584 CC Utrecht, the Netherlands}

\author{Marjolein Dijkstra}
\affiliation{Soft Condensed Matter, Debye Institute for Nanomaterials Science, Princetonplein 1, 3584 CC Utrecht, the Netherlands}

\author{Ren\'{e} van Roij}
  \affiliation{Institute for Theoretical Physics,  Utrecht University, Princetonplein 5, 3584 CC Utrecht, the Netherlands}

\date{\today}

\begin{abstract}
The formally \textit{exact} framework of equilibrium Density Functional Theory (DFT) is capable of simultaneously and consistently describing thermodynamic and structural properties of interacting many-body systems in arbitrary external potentials. In practice, however, DFT hinges on \textit{approximate} (free-)energy functionals from which density profiles (and hence the thermodynamic potential) follow via an Euler-Lagrange equation. Here we explore a relatively simple Machine Learning (ML) approach to improve the standard mean-field approximation of the excess Helmholtz free-energy functional of a 3D Lennard-Jones system at a supercritical temperature. The learning set consists of density profiles from grand-canonical Monte Carlo simulations of this system at varying  chemical potentials and external potentials in a \textit{planar} geometry only. Using the DFT formalism we nevertheless can extract not only very accurate 3D \textit{bulk} equations of state but also \textit{radial} distribution functions using the Percus test-particle method. Unfortunately, our ML approach did not provide very reliable Ornstein-Zernike direct correlation functions for small distances.  
\end{abstract}

\maketitle

\section{Introduction}

Given the massive present-day availability of computer power and data, the grown general interest in machine learning (ML) should not come as a big surprise. This interest also extends to physics,  whose community excels at  gathering, organizing, and analyzing data in order to predict and model the behavior of systems with many degrees of freedom, which is also one of the strengths of ML. An important distinction between the field of physics and ML is that physicists tend to understand, model, and predict the systems of their interest via a stepwise chain of reasoning from cause to effect, whereas ML algorithms tend to ``only'' directly relate cause to effect without necessarily understanding (in the traditional ``human'' sense) the steps in between. In other words, ML can often be regarded as a black box that is as incomprehensible as the initial raw data itself. 

Here we will also suffer, at least to some extent, from this black-box character of ML applied to a problem in classical Density Functional Theory (DFT)  \cite{evans1979nature,hansen2013theory}. However, only in a limited way because we can build on the foundations of physics to exploit, in this case, a few ingredients of the DFT formalism. As we will explain in full detail below, DFT is an exact framework to describe thermodynamic and structural properties of interacting many-body systems. This involves the solution of Euler-Lagrange equations for the equilibrium density profile for a particle system in an external potential. Now, DFT hinges for given particle-particle interactions on approximate free-energy density functionals. By comparison with Monte Carlo simulations of density profiles in a learning set of external potentials, a free-energy functional can be constructed during an ML process. The additional physics that can be extracted beyond the learning set not only includes density profiles for external potentials outside the learning set but also (i) thermodynamic bulk quantities (because the minimal value of the functional corresponds to the thermodynamic potential at equilibrium from which for instance the bulk pressure follows) and (ii) the two-body direct correlation function (because it is related to the second functional derivative of the functional) from which the radial distribution function follows. Moreover, thermodynamic surface properties such as the adsorption and the interfacial tension can be extracted from the functional. Our work is strongly inspired by recent ML work to construct a classical DFT for the Lennard-Jones (LJ) fluid in one spatial dimension  \cite{lin2018classical,lin2020analytical}, which we here extend to the three-dimensional LJ fluid. 
Similar to Refs.~\onlinecite{lin2018classical,lin2020analytical} we
use grand-canonical Monte Carlo (MC) simulations at a learning set of chemical potentials and external potentials, however in a planar geometry. We stress that the planar geometry yields an effective 1D problem embedded in 3D, not to be confused with an actual 1D problem. We will show the ability
to ``learn'' a free-energy functional that predicts density profiles of this system at chemical and external potentials {\sl outside} the training set, but also 
to extract several system properties that were {\sl not at all} present in the data of the training set, or at least not explicitly. 
In particular, we will show that from a learning set in a \textit{planar} geometry a machine-learned functional can be constructed that is capable of predicting the 3D mechanical bulk equation of state of the \textit{homogeneous} fluid (the pressure-density-chemical potential relations), the 3D \textit{radially symmetric} direct correlation function and the radial distribution function at any density, and (in principle) the prediction of Lennard-Jones density profiles in an arbitrary external potential in 3D. The agreement of these predictions against simulations varies from  very good (equation of state, radial distributions from the Percus test particle method, density profiles outside learning set) to, admittedly, rather poor (direct correlation function). The poor prediction for the latter is probably due to the rather simple form (and in retrospect perhaps an overly-simple form compared to Ref.~\onlinecite{lin2020analytical}) for the free-energy functional, and due to the treatment of the repulsive part of the LJ interaction. The main thrust of our findings at this point, therefore, is not the construction of the Lennard-Jones free-energy functional that compares ``best'' with MC simulations, but rather the notion that free-energy functionals for 3D systems can be constructed from relatively simple geometries (here planar) in the learning set. Extensions to other systems, for instance electrolytes and ionic liquids forming an electric double layer in contact with planar electrodes, could be a next step with actual applications in modelling the osmotic equation of state, the differential capacitance, and the adsorption in porous geometries. 

This paper is organised as follows. We start in section II with an extensive introduction into classical DFT -that can easily be skipped by readers familiar with this framework. In section III and IV we discuss the system and the (simulation and machine-learning) methods that we use, and in section V we discuss the resulting kernels, density profiles, equations of state, and pair correlation functions.  We end in section VI with a discussion and outlook.

\section{classical density functional theory}\label{ch:cDFT}
\subsection{Formalism}
We consider a classical one-component system of $N$ spherical particles with linear momenta ${\bf p}_i$ and center-of-mass positions ${\bf r}_i$ with $i=1,\cdots, N$ the particle label. The particles interact with each other via an isotropic pair potential $\phi(r_{ij})$, where $r_{ij}=|{\bf r}_i-{\bf r}_j|$ is the  distance between particle $i$ and $j$. All particles are  subject to a static external potential $V_{ext}({\bf r}_i)$, such that the Hamiltonian of the system reads
 \begin{align}\label{ham}
    H_N = \sum_{i=1}^N \frac{\textbf{p}^2_i}{2m} + \sum_{i<j}^N\phi(r_{ij}) + \sum_{i=1}^NV_{ext}(\textbf{r}_i),
\end{align}
where  $m$ denotes the mass of the particles. Here we note that Eq.~\eqref{ham} can also describe macroscopic bulk systems  by considering the external potential to be zero in a box of volume $V$ at a homogeneous density $\rho=N/V$ and temperature $T$. 
For these homogeneous systems, typical thermodynamic equilibrium quantities of interest include the caloric and mechanical equations of state $u(\rho,T)$ and $p(\rho,T)$ for the internal energy and pressure, respectively. Also structural quantities such as the radial distribution function $g(r)$ (at particle-particle separation $r$) and the structure factor $S(q)$ (at wavenumber $q$) are of interest for homogeneous systems \cite{hansen2013theory}.  Equilibrium statistical mechanics offers a variety of techniques to calculate (approximations to) these quantities, for instance systematic low-$\rho$ or high-$T$ expansions, integral equations based on the Ornstein-Zernike equation, or computer simulations. However, the situation is more complicated in the case of a nontrivial external potential due to, for instance, the Earth's gravity, an attractive or repulsive substrate, or a porous matrix that may confine the particles of interest. In this case, the system described by Eq.~\eqref{ham} becomes heterogeneous in thermodynamic equilibrium, such that the local density $\rho({\bf r})$ varies in space. Consequently, the energy density $u$ and the pressure $p$ become ill-defined (except, of course, within a local density approximation), and the broken translation invariance causes the radial distribution function to be of the form g(${\bf r},{\bf r}')$ rather than $g(|{\bf r}-{\bf r}'|)$. Nevertheless, the formalism of Density Functional Theory (DFT) can  provide a consistent picture of the thermodynamic and structural properties of inhomogeneous fluids in an external potential. Although DFT finds its roots in the quantum-many body description of electrons, it has also found many applications in the (essentially) classical context of soft-matter systems to describe molecular liquids, electrolytes, colloidal dispersions, etc.  \cite{evans1979nature,henderson1992fundamentals,rosenfeld1989free,hartel2012tension,hartel2015fundamental,marechal2013density,schmidt2000density,roth2010fundamental,hansen2006density}. 

DFT is  essentially a grand-canonical framework in which the temperature $T$ and the chemical potential $\mu$ of the particles are fixed to characterise the heat bath and the particle bath with which the system is in thermal and diffusive equilibrium. The corresponding thermodynamic potential is the grand potential $\Omega_0$ defined by $\beta\Omega_{0}=-\ln \sum_{N=0}^{\infty}\int d{\bf p}^N d{\bf r}^N\exp[\beta (\mu N- H_N)]/N!h^{3N}$, where $\beta^{-1}=k_BT$, $k_B$ the Boltzmann constant, and $h$ an arbitrary constant with the same dimension as the Planck constant. From $\Omega_{0}$ essentially all thermodynamic properties would follow, for instance the pressure of the homogeneous system equals $-\Omega_{0}/V$ and the internal energy is $\partial \beta\Omega_{0}/\partial\beta$. Of course, this involves the immense problem of evaluating the  $6N$-dimensional phase-space integral in the definition of $\Omega_{0}$. The key of classical DFT is that it circumvents this high-dimensional phase-space integral by a proof \cite{evans1979nature} of the existence of a grand-potential \textit{functional} $\Omega[\rho]$ of the variational one-body density profile $\rho({\bf r})$, with the properties that (i) the equilibrium density profile $\rho_0({\bf r})$ minimizes the functional $\Omega[\rho]$, and (ii) this minimum equals the equilibrium grand potential $\Omega_0$. This implies that
\begin{align}\label{eq:introductionCDFT1}
    \left.\frac{\delta\Omega[\rho]}{\delta\rho(\textbf{r})}\right|_{\rho_{0}(\textbf{r})} = 0; \quad 
    \Omega[\rho_{0}] = \Omega_0.
\end{align}
The problem is thus reduced to finding the functional $\Omega[\rho]$, and after that to solve the 3D  Euler-Lagrange equation \eqref{eq:introductionCDFT1}, which amounts to a huge reduction of the problem compared to the high-dimensional phase-space integral.

One can also prove rigorously \cite{evans1979nature,hansen2013theory,mermin1965thermal} that 
the grand potential functional $\Omega[\rho]$ can always be written as 
\begin{align}\label{eq:introductionCDFT2}
    \Omega[\rho] = \mathcal{F}[\rho] - \int \dr \rho(\textbf{r})\left(\mu - V_{ext}(\textbf{r})\right),
\end{align}
where $\mathcal{F}[\rho]$ is the \textit{intrinsic} Helmholtz free-energy functional that, and this is crucial for our machine-learning approach, \textit{only and uniquely} depends on the particle-particle interactions (here the pair potential $\phi(r)$) and on the temperature, and  \textit{not} on $\mu$ and $V_{ext}({\bf r})$. In other words, the same and unique functional $\mathcal{F}[\rho]$ for a given $\phi(r)$ applies at any chemical and external potential. That $\mathcal{F}[\rho]$ is a \textit{Helmholtz} free-energy functional follows straightforwardly from the thermodynamic relation $\Omega_0=F_0-\mu N_0$ with $N_0=\int d{\bf r}\rho_0({\bf r})$ the equilibrium number of particles and $F_0$ the equilibrium Helmholtz free energy, which can be decomposed into the sum of the potential energy 
$\int\dr\rho_0(\textbf{r})V_{ext}(\textbf{r})$ due to the external field and the remaining intrinsic free energy $\mathcal{F}[\rho_0]$.

Unfortunately, $\mathcal{F}[\rho]$ is not explicitly known in most cases. An exception is the ideal-gas case of $\phi(r)\equiv0$, where it is possible to construct the intrinsic free-energy functional as $\mathcal{F}^{id}[\rho] = k_BT\int\dr\rho(\ra)\left(\ln{\rho(\ra)\Lambda^3}-1\right)$, with $\Lambda=h/\sqrt{2\pi mk_BT}$ the thermal wavelength. The common practise in DFT  is now to split the intrinsic free energy into the ideal and the excess-over-ideal part, $\mathcal{F}[\rho] = \mathcal{F}^{id}[\rho] + \mathcal{F}^{exc}[\rho]$, and to find an explicit (usually approximate) expression for $\mathcal{F}^{exc}[\rho]$. Once such an explicit expression has been found, we can cast the minimum condition for $\rho_0({\bf r})$ of Eq.~\eqref{eq:introductionCDFT1} in the explicit form
\begin{align}\label{eq:selfconsistingequationintroduction}
    \rho_0(\textbf{r}) =\frac{\exp(\beta\mu)}{\Lambda^3}\exp{\left( - \beta\left.\frac{\delta\mathcal{F}^{exc}[\rho]}{\delta\rho(\textbf{r})}\right|_{\rho = \rho_0} -\beta V_{ext}(\textbf{r})\right)}.
\end{align}
Note that Eq.~\eqref{eq:selfconsistingequationintroduction} is a self-consistency relation for interacting systems, that usually takes the form of a nonlinear integro-differential equation that needs to be solved numerically for a given $\mu$ and $V_{ext}({\bf r})$ for a system of interest with pair potential $\phi(r)$ at temperature $T$ --- and hence with a given excess functional $\mathcal{F}^{exc}[\rho]$. In relatively simple geometries, for instance with planar or radial symmetry, a numerical solution of Eq.~\eqref{eq:selfconsistingequationintroduction} can be found at relatively low computational cost by means of e.g. a Picard iteration scheme. 

Thus, the remaining problem of DFT lies in constructing an explicit form for $\mathcal{F}^{exc}[\rho]$, for which no universal recipe is available -- not unlike the case of partition functions of interacting systems. 
There is, however, one more exact relation that can be and has been exploited, and involves the second functional derivative $-\beta\delta^2\mathcal{F}^{exc}[\rho]/\delta\rho({\bf r})\delta\rho({\bf r}')$, which equals by definition the Ornstein-Zernike direct correlation function $c({\bf r},{\bf r}')$ and is hence directly related to the two-body structure of the system. In particular, in a homogeneous bulk system the direct correlation function is of the form $c_b(|{\bf r}-{\bf r}'|)$ and its Fourier transform $\hat{c}_b(q)$ yields the structure factor $S(q)=(1-\rho\hat{c}_b(q))^{-1}$, from which the radial distribution function $g(r)$ follows by an inverse Fourier transformation. 

In this manuscript we will focus on a Lennard-Jones fluid. In the DFT treatment, we split the pair potential  $\phi(r)=\phi_0(r)+\phi_1(r)$ into a steep repulsion $\phi_0(r)$ at short distances and an attractive tail $\phi_1(r)$ of well depth $-\epsilon<0$ in accordance with Barker-Henderson theory, as further detailed in section~\ref{ch:system}. On the basis of the vast body of knowledge on the thermodynamics and the two-body structure of the hard-sphere system, extremely accurate approximations have been constructed for its intrinsic excess Helmholtz free-energy functional $\mathcal{F}_{HS}^{exc}[\rho]$, for which we will use the White-Bear mark II\cite{hansen2006density} version of the fundamental measure theory\cite{rosenfeld1989free,roth2010fundamental} throughout this paper. It is common practise in liquid-state theory to treat the attractions as a perturbation on the hard-sphere system, and a popular version results in the Van der Waals-like mean-field (MF) approximation  
\begin{align}\label{eq:MFsplitting}
    \mathcal{F}^{exc}_{MF}[\rho] \approx \mathcal{F}_{HS}^{exc}[\rho] + \frac{1}{2}\int\dr\dr'\rho(\textbf{r})\rho(\textbf{r}')\phi_1(|\textbf{r}-\textbf{r}'|).
\end{align}
The high-temperature limit of Eq.~\eqref{eq:MFsplitting} returns the hard-sphere limit, but the MF approximation fails to give accurate results for lower temperatures where the attractions play a more prominent role \cite{hansen2013theory, lin2018classical, bachelorthesis}. 
We therefore seek improved excess free-energy functionals in terms of corrections to the mean-field functional of Eq.~\eqref{eq:MFsplitting} of the quadratic and cubic form  
\begin{align}\label{eq:nosymmetryfunctional}
    \beta\mathcal{F}_{ML2}^{exc} =\;& \beta\mathcal{F}^{exc}_{MF}+ \frac{1}{2}\int \dr \dr'\rho(\textbf{r})\rho(\textbf{r}')\Omega_{2}(|\textbf{r}-\textbf{r}'|), \\
    \beta\mathcal{F}_{ML3}^{exc} =\;&  \beta\mathcal{F}_{ML2}^{exc} + \frac{1}{3}\int \dr \dr'
    \rho^2(\mathbf{r})\rho(\mathbf{r}')\Omega_3(|\mathbf{r}-\mathbf{r}'|), \label{eq:nosymmetryfunctionalML3}
\end{align}
where the labels ML2 and ML3 refer to the fact that we will use machine learning (ML) to find the optimal form of the kernels $\Omega_2(r)$ and $\Omega_3(r)$. We note that ML2 reduces to the mean-field form for $\Omega_2(r)\equiv0$ and that ML3 reduces to ML2 for $\Omega_3(r)\equiv0$.  We also emphasize that the ML3 form of the functional is \textit{not} compatible with the third-order virial-type expansion which would have entailed an additional spatial integration (say over ${\bf r}''$) and a kernel of the triple product form $f(|{\bf r}-{\bf r}'|)f(|{\bf r}'-{\bf r}''|)f(|{\bf r}''-{\bf r}|)$; finding the optimal kernel $f(r)$ proved to be computationally too demanding and inconvenient at the exploration phase of this project and hence we settled for the simpler form of the cubic term ML3.   

The building of ML functionals upon the MF functional is important. For long-ranged potentials the mean-field functional retrieves the correct asymptotic decay of the direct correlation function, such that the range of the ML corrections can conveniently be limited. For short-ranged potentials, the direct correlations are short-ranged anyway, and the inclusion of the mean-field term puts no constraint on the resulting ML functional.  

\subsection{Planar geometry}
Interestingly, we can exploit the fact that the optimal kernels $\Omega_2(r)$ and $\Omega_3(r)$ that we seek \textit{must}  be independent of $\mu$ and $V_{ext}({\bf r})$ to determine them in systems with a planar geometry, i.e. systems that have translation invariance in the $x$- and $y$-direction with external potentials $V_{ext}(z)$ and density profiles $\rho(z)$ that only depend on the normal coordinate $z$. One easily checks that the non-HS part of the MF functional of Eq.~\eqref{eq:MFsplitting} then reduces to
\begin{align}\label{eq:MFattractivepart1D}
    \mathcal{F}_{MF}^{exc}[\rho(z)] = \frac{A}{2}\int \dz\dz' \rho(z)\rho(z')\phi_{1,z}(|z-z'|),
\end{align}
where $A = \int\dx\dy$ is the (macroscopically large) area of the planar surface and $\phi_{1,z}(|z-z'|) =  \int\dx\dy\,\phi_1(\sqrt{(z-z')^2+x^2+y^2})$ is the laterally-integrated  pair potential $\phi_1(r)$. Likewise the non-HS contributions to the functionals of Eq.~\eqref{eq:nosymmetryfunctional} can be cast in the form 
 \begin{align}\label{eq:ansatz1}
    \beta\mathcal{F}_{ML2}^{exc} =\;&\beta\mathcal{F}_{MF}^{exc}+\frac{A}{2}\int \dz \dz'\rho(z)\rho(z')\omega_{2}(|z-z'|), \\
     \beta\mathcal{F}_{ML3}^{exc} =\;& \beta\mathcal{F}_{ML2}^{exc} + \frac{A}{3}
     \int \dz \dz'\rho^2(z)\rho(z')\omega_{3}(|z-z'|),\label{eq:ansatz2}
 \end{align}
where the laterally-integrated kernels $\omega_2$ and $\omega_3$ can be written as
\begin{align}\label{omi}
    \omega_i(z) = 2\pi\int_{|z|}^\infty\,dr\, r\,\Omega_{i}(r); \hspace{1 cm} i=2,3.
\end{align}
Interestingly, Eq.~\eqref{omi} can be inverted, such that we find
\begin{align}\label{Omi}
    \Omega_i(|\textbf{r}|) = -\frac{1}{2\pi}\left(\frac{1}{z}\frac{d\omega_i(z)}{dz}  \right)\bigg|_{z = |\textbf{r}|}.
\end{align}
In other words, once we find $\omega_2(z)$ and $\omega_3(z)$ from calculations in planar geometry, 
we can determine $\Omega_2(r)$ and $\Omega_3(r)$ from Eq.~\eqref{Omi} such that the direct correlation function follows by taking second functional derivatives of Eqs.~\eqref{eq:nosymmetryfunctional} and~\eqref{eq:nosymmetryfunctionalML3}. Hence, we  have access to thermodynamic as well as structural properties in bulk, in any geometric confinement, in any external potential, at any chemical potential, solely on the basis of input in a planar geometry. 

\section{System}\label{ch:system}
In this paper, we consider a 3D fluid in which the particles interact with a truncated and shifted Lennard-Jones (LJ) interaction given by
\begin{align}\label{LJ}
    \phi^{LJ}(r)=\begin{cases}
               \; 4\epsilon\left( \left( \frac{\sigma}{r} \right)^{12}-\left( \frac{\sigma}{r} \right)^{6}\right) + \epsilon_{cut},\quad &\text{for}\quad r\leq r_{cut};\\
               \; 0, &\text{for}\quad r> r_{cut},\\
            \end{cases}
\end{align}
where $\epsilon>0$ denotes the well depth and $\sigma$ is the LJ particle diameter. The full LJ potential is truncated at $r_{cut}=4\sigma$ and shifted upwards by $\epsilon_{cut}= 0.98 \cdot 10^{-3}\, \epsilon$ such that $\phi^{LJ}(r_{cut}) = 0$. The splitting of $\phi^{LJ}(r)$ into a  hard-sphere reference and an attractive tail in the DFT treatment is performed on the basis of the well-known Barker-Henderson theory 
\cite{barkerhenderson,yu2009novel} that leads to an effective and temperature-dependent hard-core diameter $0< d\leq \sigma$ that does not depend on the bulk density as explained in Refs.~\onlinecite{barkerhenderson, derivationeffdiameter}. At the temperature $k_BT/\epsilon=2$ of our main interest the effective diameter is given by $d=0.9568\sigma$. The resulting expression for $\phi_1(r)$ then reads
\begin{align}\label{LJ1}
    \phi_1(r)=\begin{cases}
               \; 0 , &\text{for}\quad r \leq \sigma;\\
               \phi^{LJ}(r), &\text{for}\quad r > \sigma.\\
            \end{cases}
\end{align}
We stress that the value inside the core, i.e. $\phi_1(r<d)$, is not uniquely defined\cite{Archer_2007,Ravikovitch_2001} and its value can be used as a fit parameter for better agreement between simulations and DFT. However, we choose here to set it to zero in line with previous studies on the LJ system\cite{Tang_1997,Tang_2002,Tang_2003}. 

\begin{figure}
    \centering
    \includegraphics[width=0.45\textwidth]{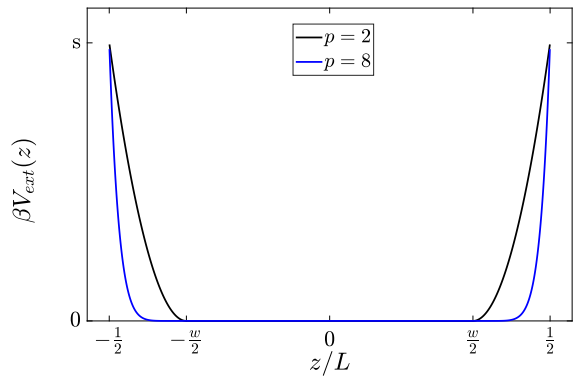}
    \caption{A general visualization of the external potential described in Eq.~\eqref{eq:Vext}. This external potential is applied in the training data and is given by the parameters $w$, $s$, $p$ and $L$. In this figure two values of $p$ are considered, namely $p=2$ (black) and $p =8$ (blue).}
    \label{fig:Vext}
\end{figure}

The external potentials $V_{ext}(z)$ that we consider in this manuscript all mimic a planar slit geometry. 
The slit is translationally invariant in the lateral $x$-$y$ plane and is mirror-symmetric in the midplane $z=0$ such that  $V_{ext}(z) = V_{ext}(-z)$.  We employ a family of external wall-particle  potentials that is repulsive and parameterised by
\begin{align}\label{eq:Vext}
    \beta V_{ext}(z) = \displaystyle \begin{cases}
                        0, \qquad\qquad         &\text{for} \quad |z|\;\leq w\frac{L}{2};\\
                        \displaystyle s\left(\frac{|z| - w\frac{L}{2}}{(1-w)\frac{L}{2}}\right)^p,  &\text{for} \quad  |z|\;> w\frac{L}{2}, 
                        \end{cases}
\end{align}
where the dimensionless strength $s=\beta V_{ext}(L/2)\geq 40$ characterises the  potential at   $|z| = L/2$, $w\in[0,1]$ denotes the width of the central part of the slit where $\beta V_{ext}(z)=0$, and $p>0$ the power that characterises the steepness of the potential.  
Fig.~\ref{fig:Vext} illustrates the external potential for general $s$, $w$, and $L$ and for steepness parameters $p=2$ (black) and $p=8$ (blue).

\section{Methods}
\subsection{Simulations}\label{ch:trainingset}
To generate the training and validation data sets to ``learn" the density functional, we perform grand-canonical Monte Carlo (MC) simulations of the 3D truncated and shifted Lennard-Jones (LJ) fluid 
confined between two planar soft-repulsive walls described by the  external potential $V_{ext}(z)$ of Eq.~\eqref{eq:Vext}.  Here, we only   consider highly repulsive walls with $s \geq 40$ to ensure that  the density reduces to essentially zero at $|z| = L/2$. 
We measure the equilibrium density profile $\rho^{MC}(z)$ in a cubic simulation box of volume $V = L^3$ with $L=10\sigma$.
We impose periodic boundary conditions in the $x$- and $y$-directions, and equilibrate the system for at least $10^5$ MC cycles before the measurements start. Each MC cycle consists of $N$ trial moves with $N$ denoting the instantaneous number of particles. Each trial move can either  displace a particle or  insert/remove a particle. The probability of selecting a trial move to displace a particle instead of an insertion/deletion move is set to 90\%. The sampling of the density profile $\rho^{MC}(z)$ is performed by dividing the volume in 320 equidistant bins that represent planar slices normal to the $z$-axis, each of width  $\sigma/32$ such that the interval $z\in[-5\sigma,5\sigma]$ is fully covered.  The density in each bin is measured and stored after every fourth MC cycle.

In order to avoid (interesting but at this stage undesired) complications due to possible phase transitions (condensation, pre-wetting, capillary evaporation, etc.), we consider only a supercritical temperature $k_BT/\epsilon=2$.
Eight different chemical potentials $\mu$ are imposed in the grand-canonical MC simulations of the LJ system, given by $\beta\mu \in \{-3.0,-2.5, ..., 0.0, 0.5\}$. 
Here, the arbitrary offset of $\mu$ is chosen such that the thermal wavelength equals the particle diameter,  $\Lambda=\sigma$; it implies that $\beta\mu\rightarrow \log\rho_b\sigma^3$ in the dilute (ideal-gas) limit $\rho_b\sigma^3\ll 1$. A total of $24$ different external potentials are considered as training sets, all with total slit length $L=10\sigma$ and strengths $s \in \{40, 60\}$, widths $w \in \{0.4, 0.65, 0.9\}$, and steepness parameters $p \in \{2, 4, 8, 10\}$. 

\begin{figure}[h!b]
    \begin{center}
    \includegraphics[width=0.48\textwidth]{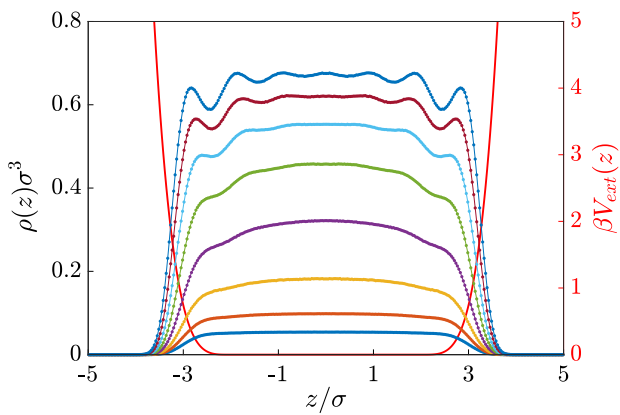}
    \end{center}
    \caption{Equilibrium density profiles $\rho(z)$ (symbols) of  a LJ fluid at temperature $k_BT/\epsilon = 2$ and  chemical potentials $\beta\mu = -3.0, -2.5, -2, -1.5, -1.0, -0.5, 0, 0.5$ from bottom to top, in an external potential $V_{ext}(z)$ (red solid line, right vertical axis) characterised by a strength $s=60$, a width $w=0.4$, and steepness parameter $p=4$  as obtained from Monte Carlo simulations.}
    \label{fig:RandomDensityProfile}
\end{figure}

As an illustration, we show in Fig. \ref{fig:RandomDensityProfile} the simulated density profiles $\rho^{MC}(z)$ of a LJ fluid at $k_BT/\epsilon=2$ and chemical potentials $\beta \mu=\{-3.0, -2.5, -2, -1.5, -1.0, -0.5, 0, 0.5\}$ (symbols) corresponding to (separately simulated) bulk densities ${\rho_b\sigma^3\approx\{0.056,0.10,0.19,0.33,0.47,0.56,0.62,0.67\}}$ 
in the external potential $V_{ext}(z)$ characterised by $s=60$, $w=0.4$, and $p=4$ as denoted by the red solid line. We observe monotonous density profiles at the lowest $\mu$'s, the development of density oscillations at higher $\mu$'s, and  a fairly well-defined  ``bulk" density in the vicinity of $z=0$ (except at the highest $\mu$'s, where the profiles of the two walls show some overlap due to the limited system size).

\subsection{Machine Learning Methods}\label{ch:OptimizationMethod}
With an optimization process that uses several techniques from the field of ML, we will construct  intrinsic free-energy functionals of the form of Eqs.~\eqref{eq:ansatz1} and~\eqref{eq:ansatz2} such that the density profiles $\rho^{ML}(z)$ that follow from this machine-learned functional are an ``optimal'' approximation to the corresponding MC densities $\rho^{MC}(z)$. We recall that $\rho^{ML}(z)$ is to be determined as a solution of the Euler-Lagrange equation \eqref{eq:selfconsistingequationintroduction}, not only for $\Lambda=\sigma$ at a given temperature, chemical potential, and external potential, but also for a given excess functional ${\cal F}^{exc}[\rho]$.
In other words, we are interested in optimal kernels $\omega_2(z)$ and $\omega_3(z)$ (where $\omega_3(z)\equiv 0$ for ML2).  

In order to quantify ``optimal'' we define a so-called loss function ${\cal L}$ that characterises the difference between ML and MC profiles, and that we will minimize with respect to $\omega_2(z)$ and $\omega_3(z)$. Here we define ${\cal L}={\cal L}_1+{\cal L}_2$ to consist of a dominant contribution ${\cal L}_1$ and a regularization term  ${\cal L}_2$. The dominant loss term is defined by the mean-square error \cite{ruehle2020data} between the MC and ML profiles,
\begin{align}\label{eq:lossfunction1}
    {\cal L}_1 = &\frac{1}{n} \sum_{j=1}^n \frac{1}{L}\int_{-L/2}^{L/2}\dz\left(\frac{\rho^{MC}_j(z) -  \rho^{ML}_j(z)}{\rho^{MC}_b(\mu_j)}\right)^2,
    \end{align}
where $j=1,\cdots,n$ labels the $n=24\times8=192$  combinations of 24 external potentials and the 8  chemical potentials of the training set as identified above. We normalise the difference between the MC and ML profiles by the MC bulk density at the chemical potential $\mu_j$ of training set $j$, for which we performed separate bulk simulations. This scaling promotes equal weights to high- and low-density states during the learning process. The regularization term ${\cal L}_2$ is independent of the MC and ML profiles and defined by
\begin{align}\label{eq:lossfunction2}
    {\cal L}_2 =  \frac{\lambda}{L}\int_{-L/2}^{L/2} \!\!\! dz\frac{1}{2}\left(\left(\frac{\omega_2(z)}{\sigma^2}\right)^2+\left(\frac{\omega_3(z)}{\sigma^{5}}\right)^2\right)f(z),
    \end{align}
where $f(z)$ is given by
\begin{align}
    f(z)=\begin{cases}
               1 \quad & |z|<\sigma;\\
               e^{z/\sigma-1} \quad & |z| \geq \sigma.
    \end{cases}
\end{align}
It accounts for the constraint that $\omega_2(z)$ and $\omega_3(z)$ must decay smoothly to zero for $z\gg\sigma$, where our statistics is poor. Moreover, ${\cal L}_2$ also suppresses undue high-wavenumber undulations that tend to develop at $|z|<\sigma$. We tune the (positive) regularization parameter $\lambda$ by trial and error such that it contributes less to the minimization procedure than $\mathcal{L}_1$, 
while not being too small to be irrelevant. Note that $\mathcal{L}_2$ effectively reduces the range of $\omega_i(z)$ by suppressing it exponentially for $|z|>\sigma$. 

The minimization of the total loss function ${\cal L}$ is performed with the stochastic and iterative optimization method Adam as proposed by Kingma and Ba in Ref.~\onlinecite{kingma2014adam}. We use their suggested default step size $\alpha=0.001$ and exponential decay rates $\beta_1=0.9$, $\beta_2=0.999$, and refer the reader to their work for a full description of the method and its parameters. During each iteration of the minimization process the gradient of the loss function w.r.t. the kernels $\omega_i(z)$ is required, which are straightforwardly derived for ${\cal L}_2$ to be the functional derivatives
 \begin{align}\label{L2domi}
     \frac{\delta {\cal L}_2}{\delta \omega_i(z)}=\frac{\lambda}{L}\frac{\omega_i(z)}{\sigma^{6i-8}}f(z).
 \end{align}
The functional derivatives of ${\cal L}_1$ with respect to $\omega_i(z)$ for $i=2,3$ is more involved and stems from the dependence of ${\cal L}_1$ on the ML density profiles $\rho_j^{ML}(z)$ for $j=1,\cdots, n$, such that the functional chain-rule yields  
\begin{align}\label{L1domi}
    \frac{\delta {\cal L}_1}{\delta \omega_i(z)} =\; &\frac{-2}{n}\sum^n_{j=1}\frac{1}{L}\int_{-L/2}^{L/2}\dz'\,\frac{\rho_j^{MC}(z') - \rho_j^{ML}(z')}{(\rho_b^{MC}(\mu_j))^2}\frac{\delta\rho_j^{ML}(z')}{\delta\omega_i(z)},
\end{align}
where we replaced the dummy integration variable $z$ of Eq.~\eqref{eq:lossfunction1} by $z'$. From the Euler-Lagrange equation \eqref{eq:selfconsistingequationintroduction} for the DFT equilibrium profiles $\rho_0$ -- which are  represented by $\rho_j^{ML}$ in Eq.~\eqref{L1domi} -- one checks that  $\delta\rho_0(z')/\delta\omega_i(z)=-\rho_0(z')\times\delta^2 \beta {\cal F}^{exc}/\delta\rho(z')\delta\omega_i(z)$. Upon considering $\omega_i$ and $\rho$ independent variables in Eqs.~\eqref{eq:ansatz1}  and~\eqref{eq:ansatz2} for the ML2 and ML3 excess functional, respectively, the second (cross) derivative for $i=2$ equals $\rho_0(z+z')$, and for ML3 and $i=3$ it equals $\frac{1}{3}[\rho_0^2(z+z') + 2\rho_0(z')\rho_0(z+z')]$. 
Hence, within this approximation a numerical integration of $z'$ suffices to evaluate Eq.~\eqref{L1domi}, and in combination with Eqs.~\eqref{L2domi} we can numerically calculate $\delta {\cal L}/\delta \omega_i(z)$ for the grid points $z$ of our system. Thus, we have all ingredients to minimize ${\cal L}$ by means of Adam \cite{kingma2014adam}.

\subsection{The Training Process}\label{ch:trainingprocess}
The training process starts with an initial guess for the two kernels, for which we take the MF approximation $\omega_2^0(z) = \omega_3^0(z)\equiv 0$, where the superscript $0$ denotes the $0$-th iteration in the training process. Next, we use these kernels to calculate the $n$ density profiles $\rho_{j,k}^{ML}(z)$ for learning sample $j=1,\cdots, n$ and iteration label $k=0$ by solving the Euler-Lagrange equation Eq.~\eqref{eq:selfconsistingequationintroduction} using a Picard iteration scheme with the MC profile $\rho_j^{MC}(z)$ as the initial guess.

On the basis of Eqs.~\eqref{eq:lossfunction1}-\eqref{L1domi} we can then evaluate ${\cal L}$ and $\delta {\cal L}/\delta \omega_i(z)$ for $i=1,2$, from which improved kernels $\omega_i^{k}(z)$ are constructed for $k=1$ by employing Adam\cite{kingma2014adam}, which will give rise to improved density profiles $\rho_{j,1}^{ML}(z)$, etc. 
For $k\geq 2$ we take $\rho_{j,k-1}^{ML}(z)$ as initial guess in the Picard-iteration of $\rho_{j,k}^{ML}$. The iteration process is repeated until the loss function has converged.

Although Adam is already an efficient algorithm for the learning process, its computational cost can be significantly 
reduced by making use of stochastic optimization. Rather than using all $n$ elements of the training set in every iteration, which involves the addition of all $n$ terms in Eq.~\eqref{L1domi} at every iteration level $k$, we consider mini batches with only 20 randomly selected elements of the training set during each Picard iteration $k$.  
The gradient of the loss function ${\cal L}_1$ is computed by only taking into account this mini batch, thus the summation over the $n$ density profiles of Eq.~\eqref{L1domi} changes to a summation over $20$ randomly-selected density profiles and the normalisation factor $1/n$ becomes $1/20$. A new mini batch is randomly selected during every iteration in the ML process.


\section{Results for the Lennard-Jones System}\label{ch:resultsLJ}


We perform MC simulations of the LJ system to generate MC density profiles with $24$ different external potentials, described in section \ref{ch:system}, and $8$ equi-distant different chemical potentials, $\beta\mu \in \{-3.0, -2.5,\cdots, 0.0, 0.5\}$, for the temperature $k_BT/\epsilon = 2$. We describe the kernels, the resulting density profiles, the mechanical equations of state of the bulk fluid, and the  radial distribution functions that follow from the functionals ML2 and ML3 using two different routes.


\subsection{The Kernels}
For several choices of the regularisation parameter $\lambda$ in Eq.~\eqref{eq:lossfunction2} we determined the kernel $\omega_2(z)$ for ML2 and  $\omega_2(z)$ and $\omega_3(z)$ for ML3. Without a significant $\mathcal{L}_2$ contribution, $\lambda\leq 10^{-3}$, we found spurious peaks in both $\omega_2(z)$ and $\omega_3(z)$ for $|z|>8\sigma$, i.e. at the largest separations (with the poorest statistics) we considered in the learning set; these spurious peaks disappeared and $\omega_2(z)$ smoothly decayed to zero for $\lambda \geq 10^{-2}$, and throughout we settle for $\lambda=10^{-2}$ as a reasonable compromise between error-correction and minimization of the actual loss function of interest ${\cal L}_1$. 

\begin{figure}
    \centering
    \includegraphics[width=0.47\textwidth]{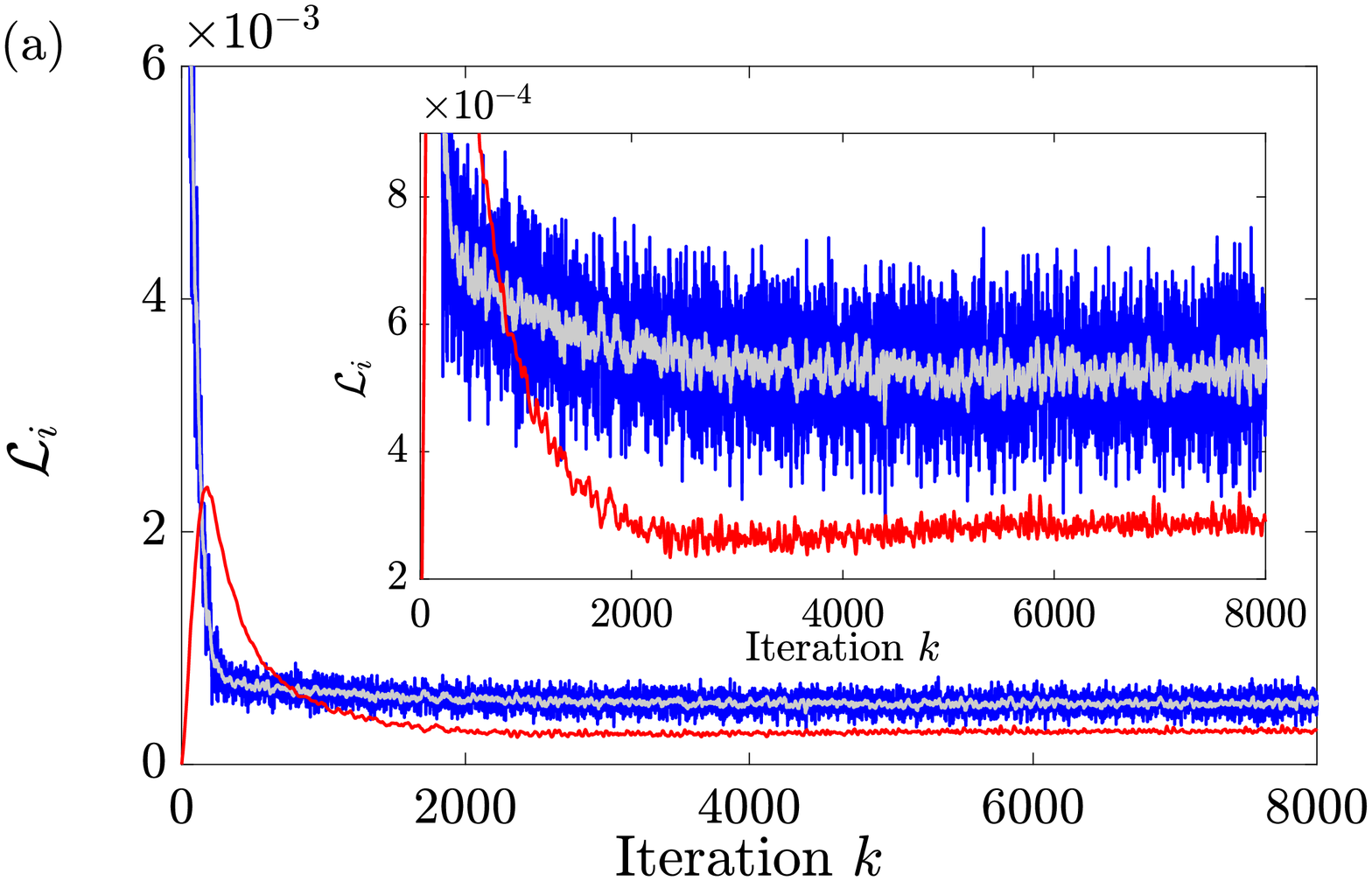}
    \includegraphics[width=0.47\textwidth]{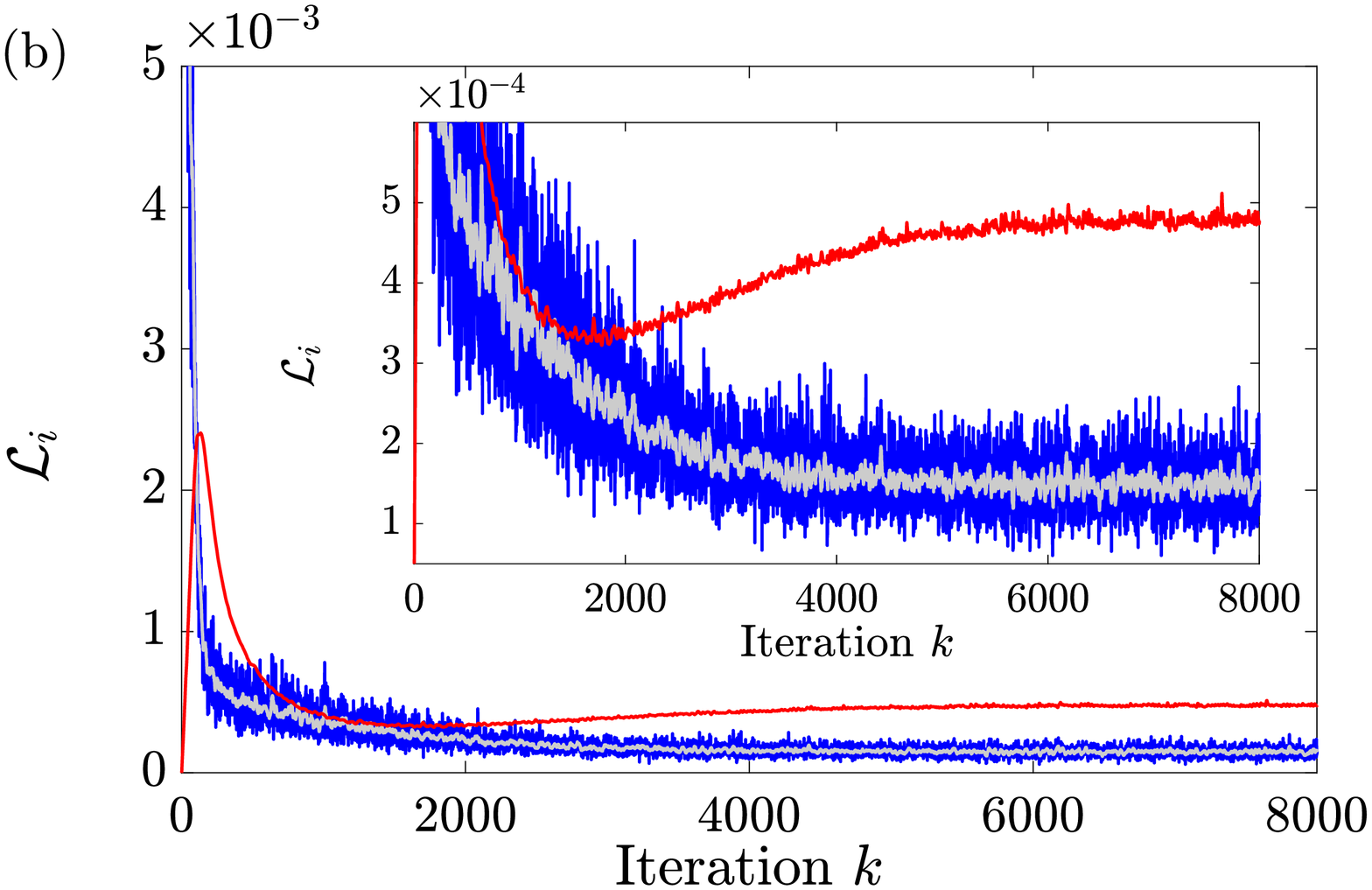}
    \caption{The loss function contributions  $\mathcal{L}_1$ (blue) and $\mathcal{L}_2$ (red) as function of the iteration label $k$, in (a) for ML2 and in (b) for ML3. The grey traces in (a) and (b) represent moving average of $\mathcal{L}_1$, and the insets only zoom in. }
    \label{fig:loss_iteration}
\end{figure}

In Fig.~\ref{fig:loss_iteration} we present the evolution of the loss functions ${\cal L}_1$ (blue) and ${\cal L}_2$ (red) during the training process with iteration label $k$, in (a) for ML2 and in (b) for ML3; the grey curves in (a) and (b) represent a moving average of $\mathcal{L}_1$ over 15 iterations. We observe good convergence after, say  $k=5000$ iterations, respectively. We note that the minimized loss function ${\cal L}_1$ of ML2 is as small as $5\times 10^{-4}$, and for ML3 it is even about four times smaller. We also note that ${\cal L}_2<{\cal L}_1$ for ML2, as desired for a regularisation term that is (naively) supposed to be a small correction to the total loss function. However, for ML3 we find ${\cal L}_1$ to be so small that it has dropped below the regularization term ${\cal L}_2$, which in retrospect should be seen as a consequence of the good accuracy of the ML3 functional rather than as a problem for the relative magnitude of the two contributions to the loss function. 

\begin{figure}[ht!]
    \centering
               \includegraphics[width=0.47\textwidth]{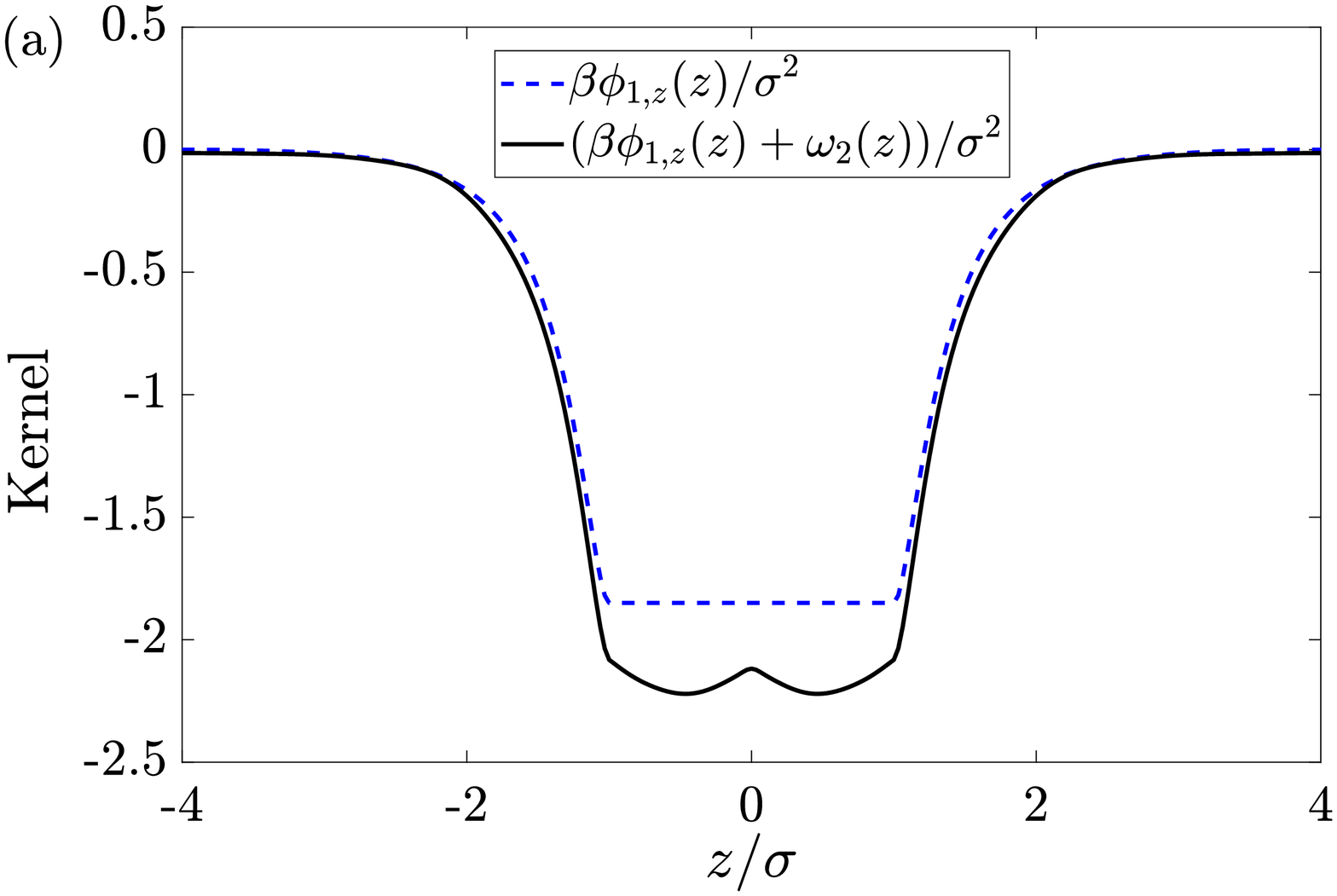}\\
       \includegraphics[width=0.47\textwidth]{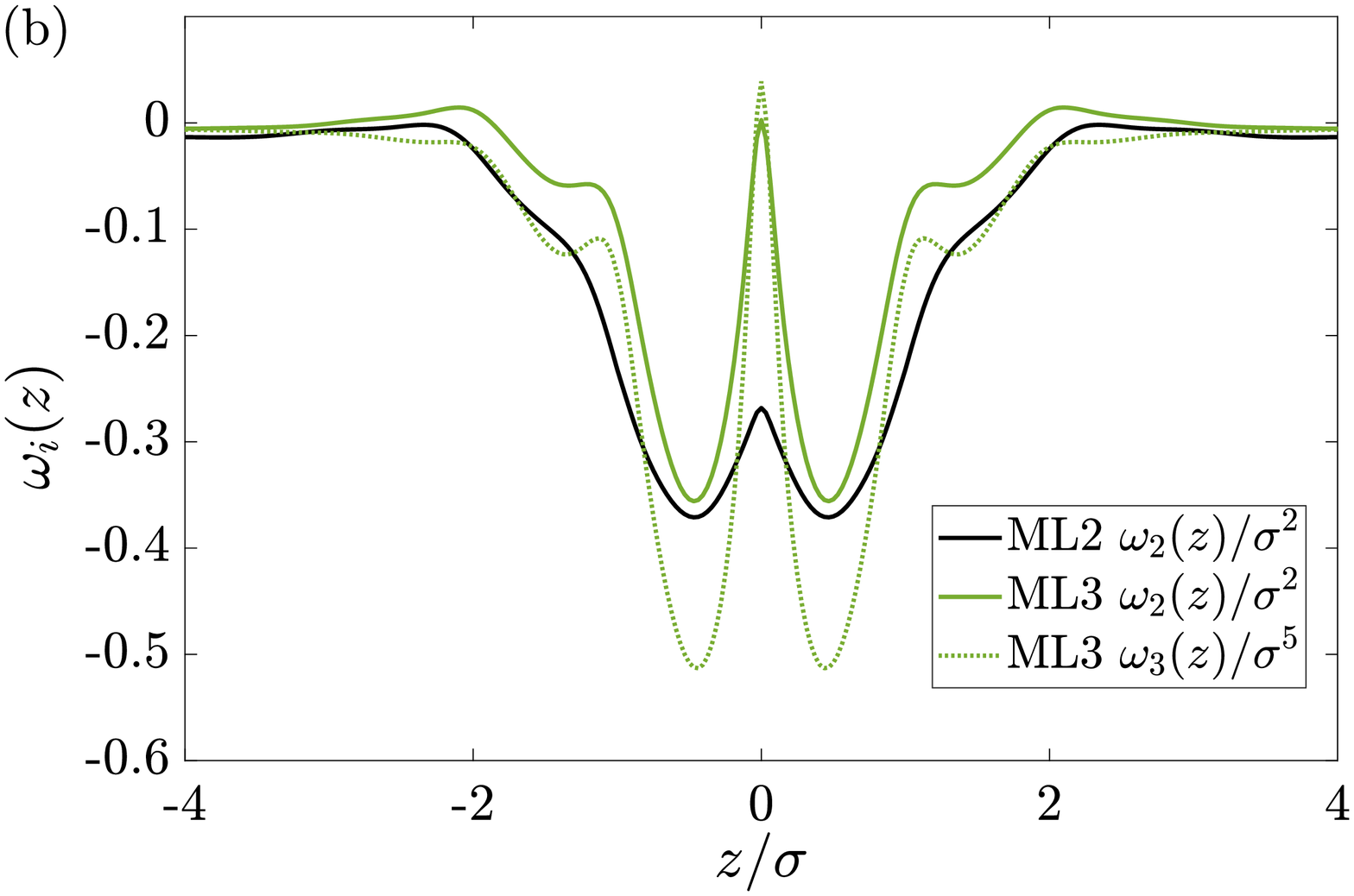}
    \caption{(a) The mean-field (MF) kernel $\beta\phi_{1,z}(z)/\sigma^2$ (dashed blue line) and its quadratic Machine Learning (ML2) improvement $(\beta\phi_{1,z}(z)+\omega_2(z))/\sigma^2$ (black solid line), as obtained for the LJ system at temperature $k_BT/\epsilon=2$.   (b) The cubic Machine Learning (ML3) kernels $\omega_2(z)/\sigma^2$ (green solid line) and $\omega_3(z)/\sigma^5$ (green dotted line), also at $k_BT/\epsilon=2$, for comparison together with the ML2 kernel $\omega_2(z)/\sigma^2$ (black solid line).}
    \label{fig:Kernel2lalleTLJ}
\end{figure}

In Fig.~\ref{fig:Kernel2lalleTLJ}(a) we show the MF (scaled) kernel $\beta\phi_{1,z}(z)$  (dashed blue line) and its ML2 correction
$\beta\phi_{1,z}(z)+\omega_2(z)$ (black solid line), as obtained after 5000 iterations. For all $z$ the ML2 kernel is more negative than the MF kernel, as if there is actually more cohesive energy in the system than predicted by MF. We see that $\omega_2(z)$ develops a peculiar and unexpected small ``bump'' close to $z=0$.  For ML3 a similar feature occurs close to $z=0$ in both $\omega_2(z)$ and $\omega_3(z)$, as can be seen in Fig. \ref{fig:Kernel2lalleTLJ}(b) where we plot $\omega_2(z)$ (green solid line) and $\omega_3(z)$ (green dotted line) for the ML3 case as obtained after 5000 iterations, together with the ML2 kernel $\omega_2(z)$ (black solid line) for comparison.  We see that $\omega_2$ from ML3 is again essentially negative (except for a tiny positive feature at $z=0$ and  $|z|\simeq2\sigma$),  and contains a ``bump'' similar to the ML2 case. We also see that $\omega_3(z)$ has a structure that is quite similar to $\omega_2(z)$, however more pronounced with a higher peaks and lower valleys.  Below we will investigate the thermodynamic and structural properties that follow from DFT based on these kernels.

\subsection{The Density Profiles}\label{ch:LJdensityprofiles}
\begin{figure}
   \includegraphics[width=0.47\textwidth]{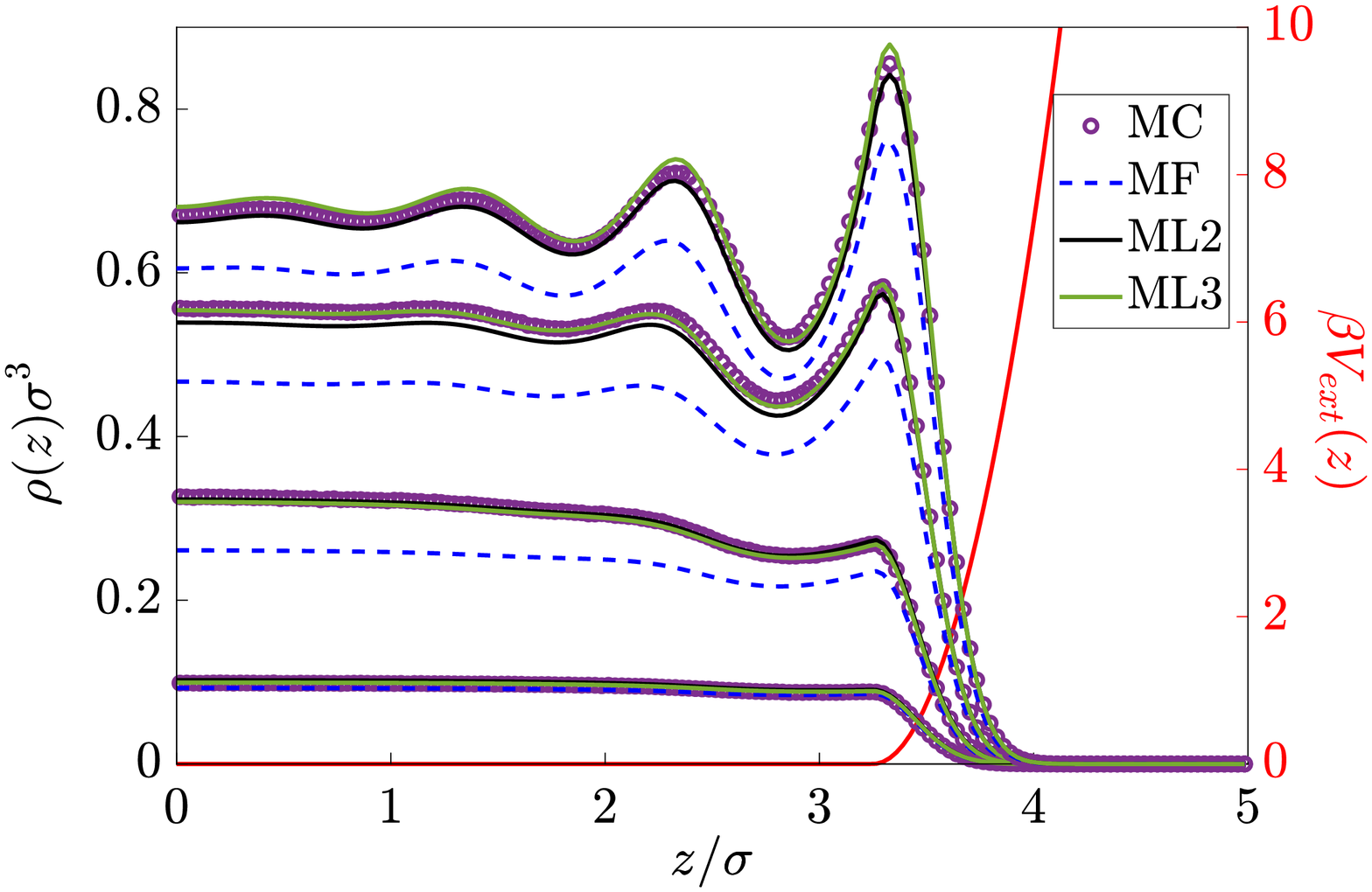}
    \caption{\label{fig:DPkernel2LJ} Density profiles of a (truncated) Lennard-Jones fluid confined in a planar slit characterised by a repulsive external potential given by Eq.~\eqref{eq:Vext} with parameters $w = 0.65,\;p = 2.0,\;s = 40$ at temperature $k_BT/\epsilon = 2$ and at four chemical potentials $\beta\mu \in \{-2.5, -1.5, -0.5, 0.5\}$ from bottom to top. Symbols stem from the grand-canonical MC simulations, and curves from the MF (blue dashed),  ML2 (black solid), and ML3 (green solid) functionals; all four state points are part of the training set.}
\end{figure}

The first test of the quality of the ML functionals is a comparison of their resulting density profiles with the simulated ones from the training set. In Fig.~\ref{fig:DPkernel2LJ} this comparison is illustrated for the external potential parameterised by $w = 0.65$, $p = 2.0$, and $s = 40$ (shown in red) and the four chemical potentials $\beta\mu \in \{-2.5, -1.5, -0.5, 0.5\}$; for symmetry reasons we only plot the regime $0<z<L/2$, and for comparison we also show the MF profiles. Clearly, the MF predictions are substantially worse than ML2 and ML3, except at the lowest $\mu$, and ML3 constitutes a small  improvement over ML2, especially at the peaks of the profiles at intermediate to high  $\mu$. 
In fact we can also conclude from Fig.~\ref{fig:DPkernel2LJ} that the main improvement of ML2 and ML3 over MF compared to the simulations concerns the bulk density $\rho_b$ that is approached in the center of the slit at $z=0$, as will be made more explicit below. 

\begin{figure*}
\includegraphics[width=0.98\textwidth]{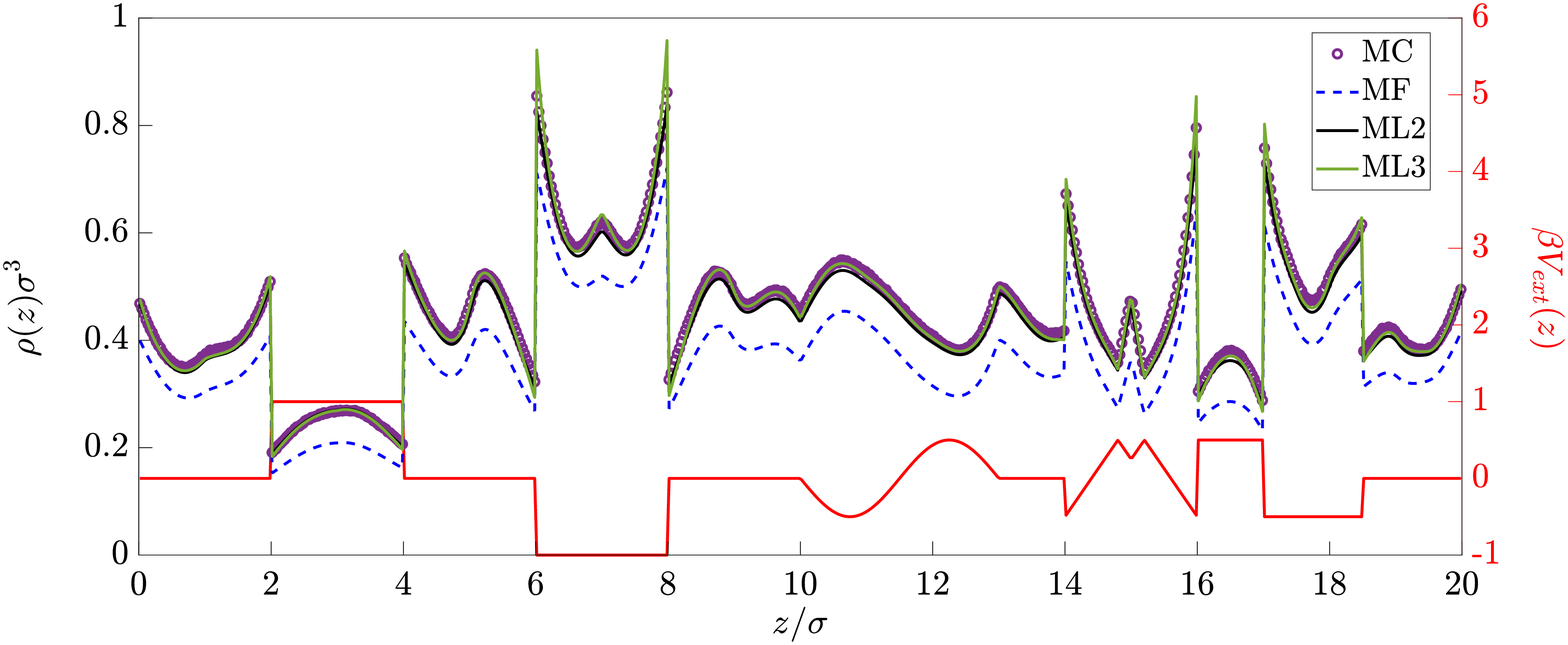}
    \caption{\label{fig:DP_newVext} The equilibrium density profile for a LJ fluid at chemical potential $\beta \mu = -1$ and temperature $k_BT/\epsilon = 2$ in the external potential $V_{ext}(z)$ (red solid line) outside the ML training set. Symbols stem from grand-canonical Monte Carlo (MC) simulations, and lines are DFT predictions based on the mean-field (MF) approximation (blue dashed) and on the quadratic (ML2, black solid) and cubic (ML3, green solid) corrections with machine-learned kernels.}
\end{figure*}

In Fig. \ref{fig:DP_newVext} we consider a comparison of MC simulations with MF, ML2, and ML3 density profiles in a particular external potential outside the training set, at $\beta\mu=-1$. The external potential consists of hard walls at $z=0$ and $z=20\sigma$, and for $z\in[0,20\sigma]$ the potential varies irregularly with wells and barriers between $\pm k_BT$ as shown by the solid red curve in Fig. \ref{fig:DP_newVext}. We see again that both ML2 and ML3 are largely of comparable quality and substantially more accurate than MF.

\subsection{Mechanical equation of state of the bulk}
\begin{figure}
        \includegraphics[width=0.47\textwidth]{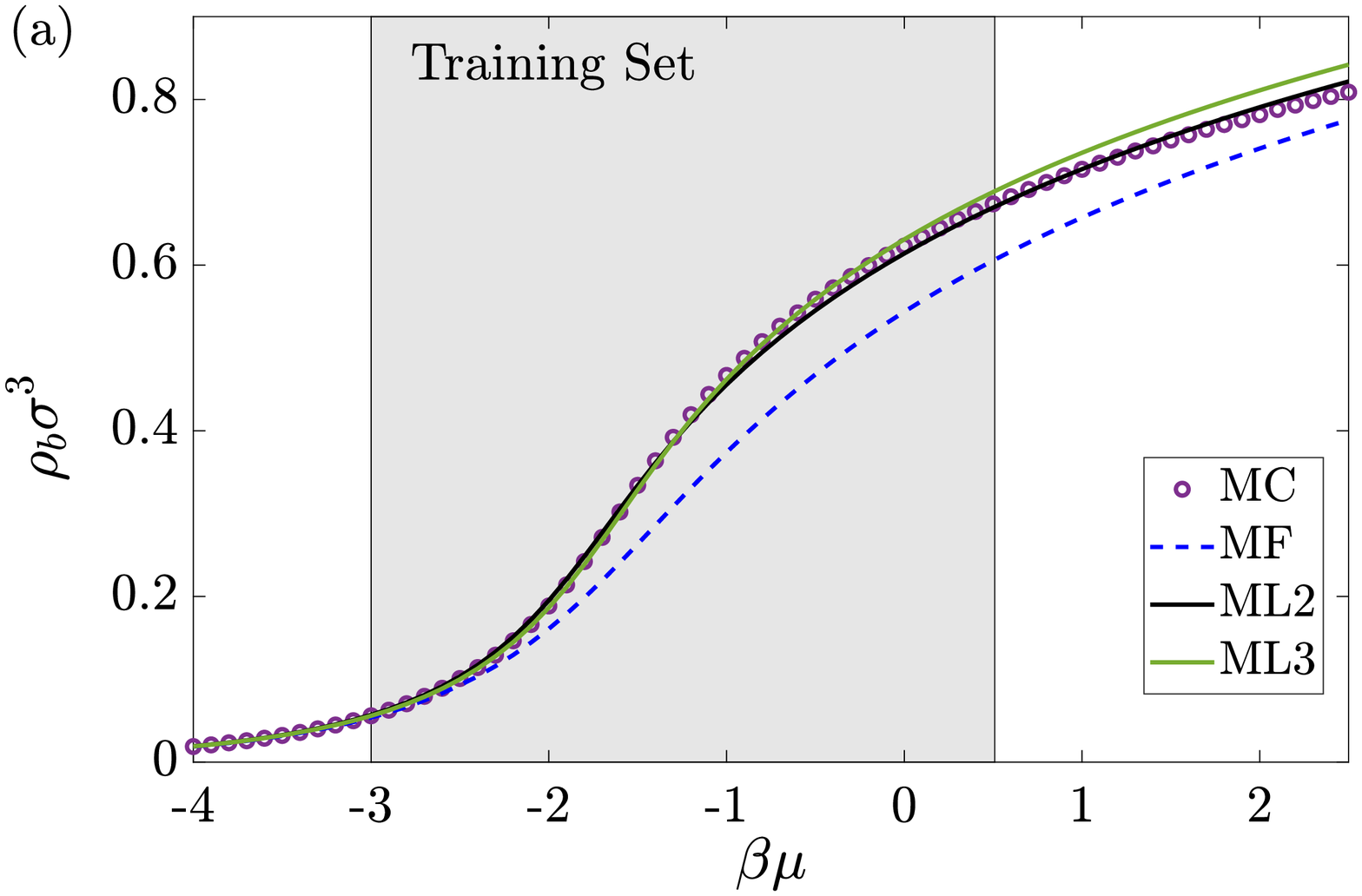} \includegraphics[width=0.47\textwidth]{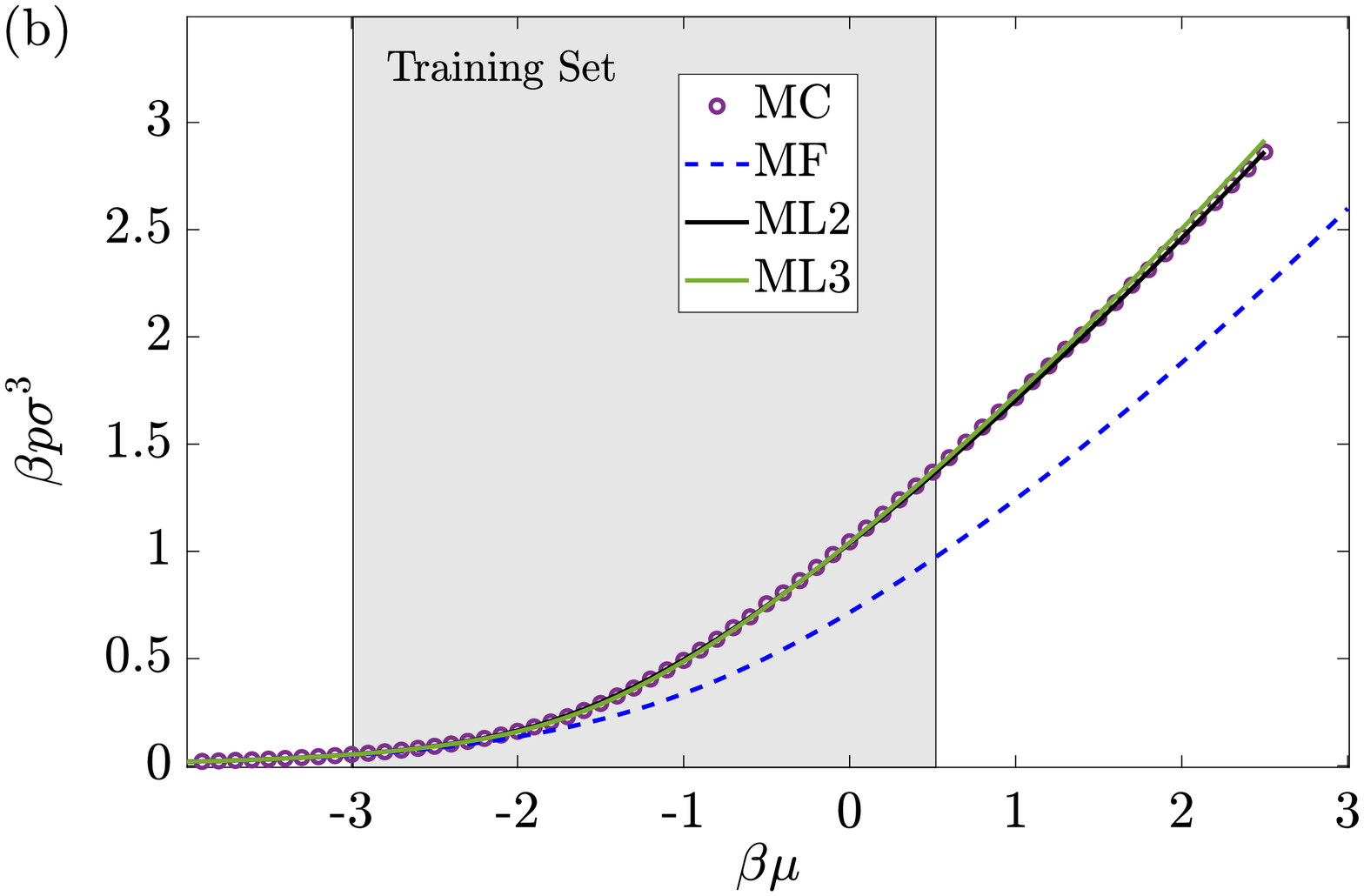}
    \includegraphics[width=0.47\textwidth]{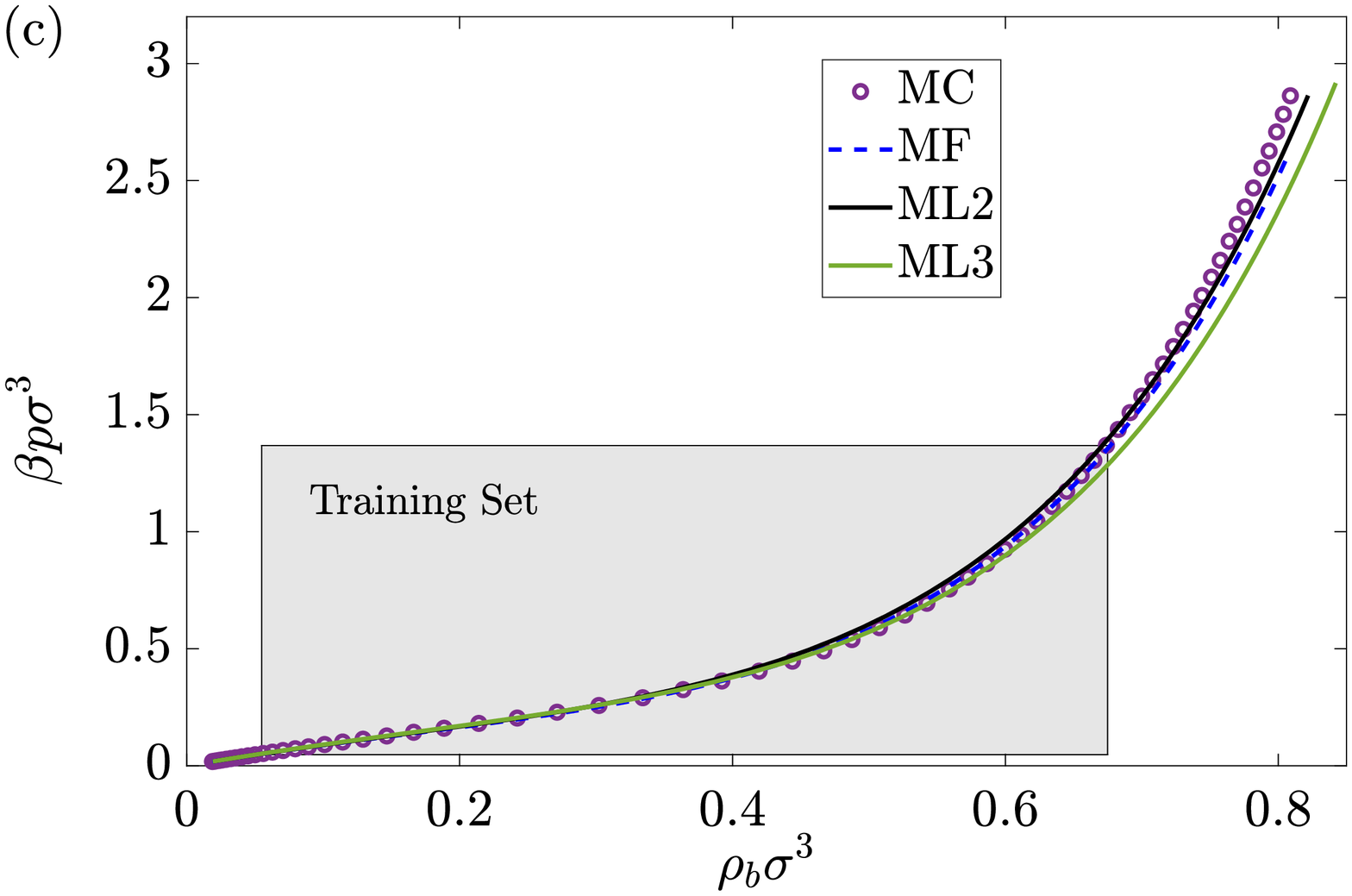}
    \caption{\label{fig:Bulkdensity_LJ}The relations between (a) the bulk density $\rho_b$ and (b) the bulk pressure $p$ as a function of the chemical potential $\mu$ for the (truncated) Lennard-Jones fluid at temperature $k_BT/\epsilon=2$ as obtained from grand-canonical Monte Carlo  simulations (MC, symbols), the machine learning functionals ML2 (black) and ML3 (green), and the mean-field function (MF, blue dashed). In (c) the corresponding pressure-density relation $p(\rho_b)$ are shown, obtained by elimination of $\mu$ from $\rho_b(\mu)$ of (a) and $p(\mu)$ of (b).
    }
\end{figure}

The (isothermal) mechanical bulk equation of state provides relations between the bulk density $\rho_b$, the pressure $p$, and the chemical potential $\mu$, satisfying the constraint of the Gibbs-Duhem equation $d\!p=\rho_b \,d\!\mu$ such that we can equivalently consider $\rho_b(\mu)$, $p(\mu)$, or $p(\rho_b)$. 
Within DFT the bulk density $\rho_b(\mu)$ that follows from a particular free-energy excess functional follows from the solution of the Euler-Lagrange equation \eqref{eq:selfconsistingequationintroduction} for the homogeneous bulk case $V_{ext}\equiv 0$, which reduces for ML2 and ML3 to a nonlinear algebraic equation with coefficients that depend on $\int dz\,\omega_i(z)$. Hence $\rho_b(\mu)$ is straightforwardly solved numerically for the the three functionals ML2, ML3, and MF of our interest here. The bulk pressure follows as $p(\mu)=-\Omega[\rho_b]/V$, from which $p(\rho_b)$ follows upon inversion of $\rho_b(\mu)$. For the temperature of interest, $k_BT/\epsilon=2$, these three representations of the equation of state
are shown in Fig. \ref{fig:Bulkdensity_LJ}(a)-(c) for the three functionals MF (blue dashed line), ML2 (black solid line), and ML3 (green solid line) together with the MC data (purple symbols). The regime of the training set is hatched grey. In the $\mu$-dependent curves of (a) $\rho_b(\mu)$ and (b) $p(\mu)$,  
we find agreement in the low-density limit $\beta\mu<-3$, as expected, since all functionals include the ideal-gas limit properly. In the regime of the training set we also see ML2 and ML3 outperforming MF by a large margin in (a) and (b), with a small but hardly noticable improvement of ML3 compared to ML2, as we could have expected on the basis of the density profiles of Fig.~\ref{fig:DP_newVext} and the loss functions of Fig.~\ref{fig:loss_iteration}. At the high-$\mu$ side outside the training set, Fig.~\ref{fig:Bulkdensity_LJ}(a) shows an increasingly deteriorating quality of the ML2 and especially the ML3 prediction, which are systematically higher than the MC data, although they are still much more accurate than the predictions based on the MF functional. Interestingly, however, the picture that emerges from the $p(\rho_b)$ representation shown in \ref{fig:Bulkdensity_LJ}(c) is much more forgiving for the MF functional, which is now of comparable good agreement in the complete regime of the training set and deviates as much as ML2 (and even less than ML3) from the MC data. Clearly, this relatively good MF and ML2 performance is due to a fortunate cancellation of errors occurring in the process of eliminating the dependence on the chemical potential. 

It is perhaps remarkable that rather accurate $bulk$ equations of state in a complete density interval can be obtained from MC simulations at only a few chemical potentials in only a few external potentials. Here it is crucial to appreciate the DFT formalism, which includes the statement that the intrinsic excess free-energy functional ${\cal F}^{exc}[\rho]$ that we construct by the ML Ans\"{a}tze of Eqs.~\eqref{eq:ansatz1} and~\eqref{eq:ansatz2} is independent of the external and chemical potential, and hence can also be applied at any $\mu$ in the homogeneous bulk where $V_{ext}\equiv 0$.

\subsection{The structure of the bulk fluid}
A key feature of DFT is that it provides not only thermodynamic but also structural information, where we have seen that the first functional derivative $\delta  {\cal F}^{exc}[\rho]/\delta\rho({\bf r})$ plays a key role in the Euler-Lagrange equation \eqref{eq:selfconsistingequationintroduction} for the equilibrium one-body distribution function.  We have also seen already that the second functional derivative 
$-\beta\delta^2\mathcal{F}^{exc}[\rho]/\delta\rho({\bf r})\delta\rho({\bf r}')\equiv c({\bf r},{\bf r}')$ equals the direct correlation function and governs the two-body distribution function\cite{ hansen2013theory,evans1979nature}. In a homogeneous and isotropic bulk fluid symmetry dictates that the direct correlation  takes the bulk form $c_b(|{\bf r}-{\bf r}'|)$, and the radial distribution function $g(r)$ follows from the Ornstein-Zernike equation $g(r)-1=c_b(r)+\rho_b\int d{\bf r'} (g(r')-1)c_b(|{\bf r}-{\bf r }')|)$. Since the ML2 and ML3 functionals have been fully determined in terms of $\omega_i(z)$ in planar geometry and its conversion to $\Omega_i(r)$ according to Eq.~\eqref{Omi}, we can write from Eqs.~\eqref{eq:ansatz1} and~\eqref{eq:ansatz2}
\begin{equation}
c_b(r)=c_{HS}(r)-\beta\phi_1(r)-\Omega_2(r) - 2\rho_b \Omega_3(r),   \label{cb} 
\end{equation}
where for $c_{HS}(r)$ we used the White-Bear mark I direct correlation function reported in Ref.~\onlinecite{Roth_2002} (and where $\Omega_3\equiv 0$ for ML2). Consequently, our ML2 and ML3 functionals (and likewise also the MF functional) give direct access to the two-body structure encoded in $c_b(r)$ and $g(r)$.

\begin{figure}
    \includegraphics[width=0.47\textwidth]{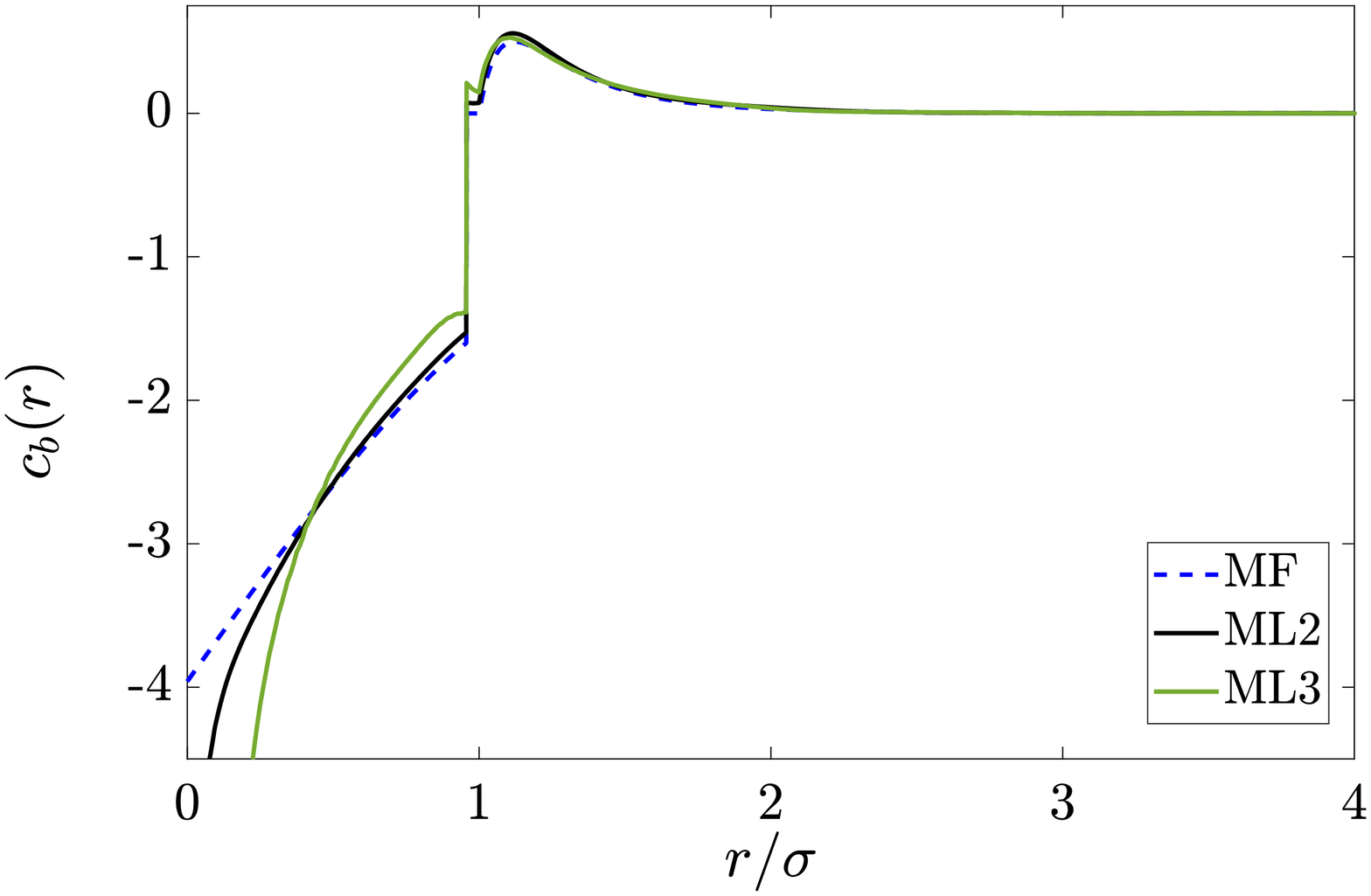}
    \caption{\label{fig:DCF_LJ} The Ornstein-Zernike direct correlation function $c_b(r)$ of the bulk Lennard-Jones system at temperature $k_BT/\epsilon=2$ and bulk density $\rho_b\sigma^3 = 0.39$, as predicted by the second functional derivative of the excess free-energy functional within the mean-field (MF) approximation and its quadratic (ML2) and cubic (ML3) Machine Learning corrections.
    }
\end{figure}

In Fig. \ref{fig:DCF_LJ} we plot the resulting $c_b(r)$ for bulk density $\rho_b\sigma^3 = 0.39$, and find fairly good agreement between MF, ML2, and ML3, except close to $r=0$ where the $c_b(r)$ from ML2 and especially ML3 become deeply negative. This can be traced back directly to Fig.~\ref{fig:Kernel2lalleTLJ}, which reveals that (i) there is close structural similarity between the functionals outside the hard core, and (ii) that the ``bumps" of $\omega_2(z)$ and $\omega_3(z)$ close to $z=0$ give rise to a substantial slope $d\!\omega_3(z)/d\!z$ for $z/\sigma\in[0,0.05]$  that translates via Eq.~\eqref{Omi} in a relatively large effect in $c_b(r)$ in the vicinity of $r=0$. Note also that $c_b(r)$ vanishes for $d<r\leq\sigma$ due to the Barker-Henderson splitting, and that all three versions of $c_b(r)$ agree pretty accurately outside the hard core, at least on the scale of the plot. 

\begin{figure}
        \includegraphics[width=0.47\textwidth]{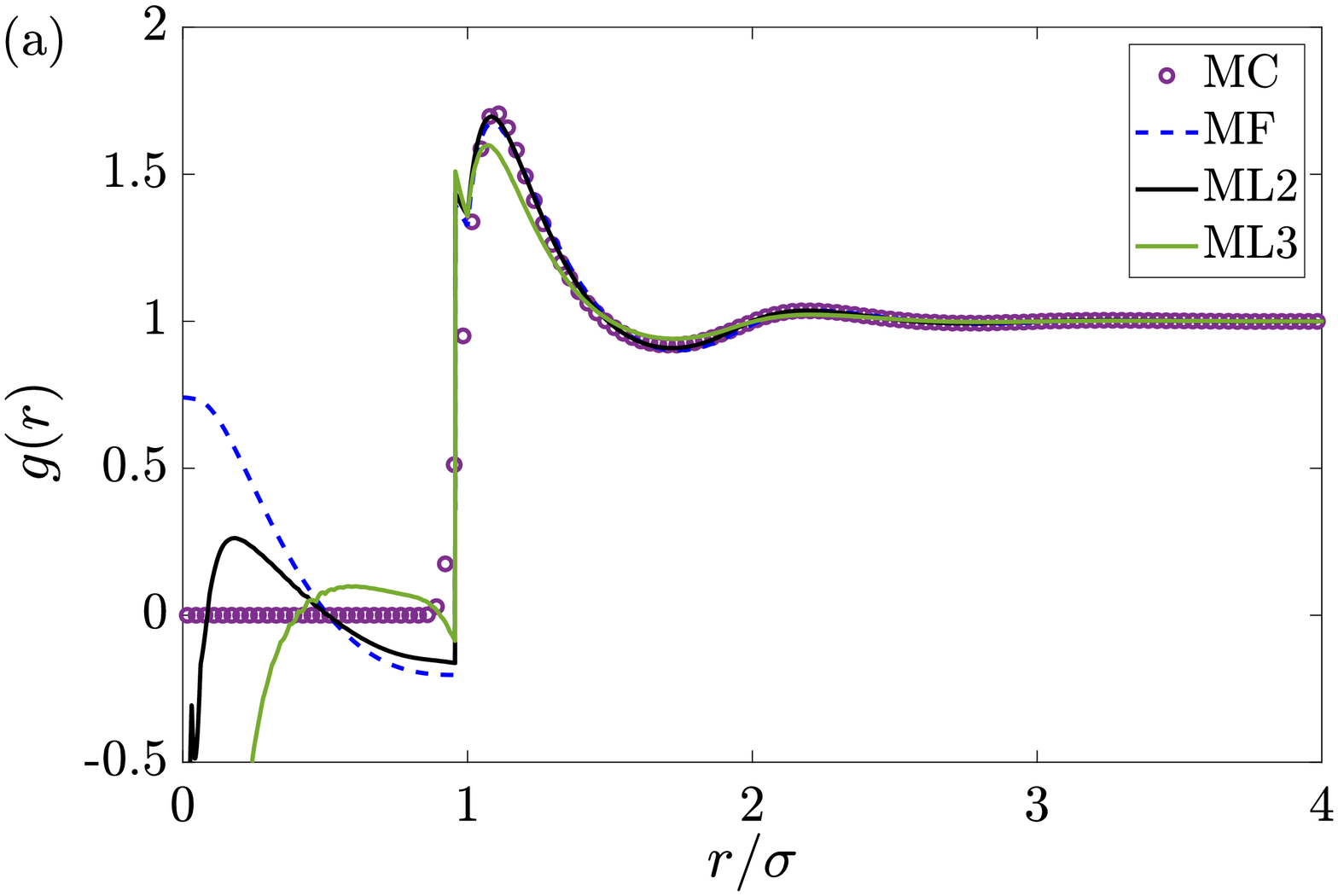}
    \includegraphics[width=0.47\textwidth]{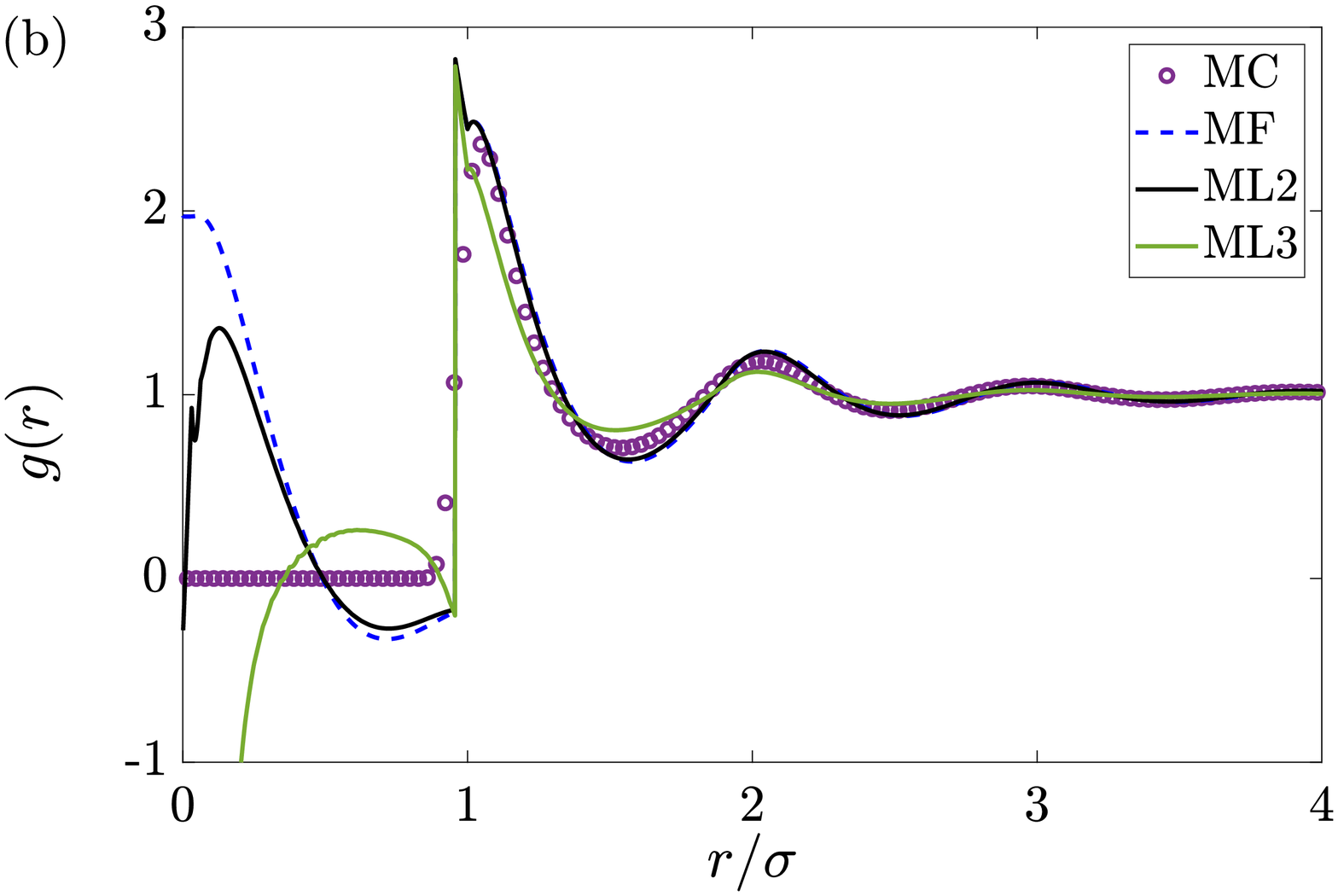}
    \caption{\label{fig:OZ_lowrho_LJ} The radial distribution function $g(r)$ of a truncated Lennard-Jones fluid at bulk density (a) $\rho_b\sigma^3 = 0.39$ and (b) $\rho_b\sigma^3 = 0.837$, as obtained from the Ornstein-Zernike equation with a direct correlation function $c_b(r)$ that follows from the free-energy functionals ML2 (black), ML3 (green), and MF (blue dashed). The symbols denote $g(r)$ as obtained from canonical Monte Carlo simulations at the same bulk density and temperature.}
\end{figure}

Upon insertion of $c_b(r)$ into the (Fourier transform of the) Ornstein-Zernike equation, we find (after an inverse Fourier transform) the radial distribution functions $g(r)$  that we compare with canonical MC simulations at a given density $\rho_b$ (at the fixed temperature of interest $k_BT/\epsilon=2)$. The three lines in Fig.~\ref{fig:OZ_lowrho_LJ}  show these radial distribution functions for MF, ML2 and ML3 at bulk density (a) $\rho_b\sigma^3 = 0.39$ and (b) $\rho_b\sigma^3 = 0.837$, together with the MC simulation results (symbols).  For both the lower density in (a) and the higher one in (b) we find  reasonably good overall agreement outside the hard core ($r>d$), with MF and ML2 actually outperforming ML3 close to contact. 
Inside the hard core our prediction for the $g(r)$ is poor in all cases, which is not surprising given that the underlying $c_{HS}(r)$ is constructed such as to cause a vanishing $g(r)$ inside the hard core; any tampering of the direct correlation (such as adding terms as we do in Eq.~\eqref{cb}) will give rise to spurious nonzero contributions to $g(r)$ for $r<d$\cite{Archer_2017}.

\begin{figure}
    \includegraphics[width=0.47\textwidth]{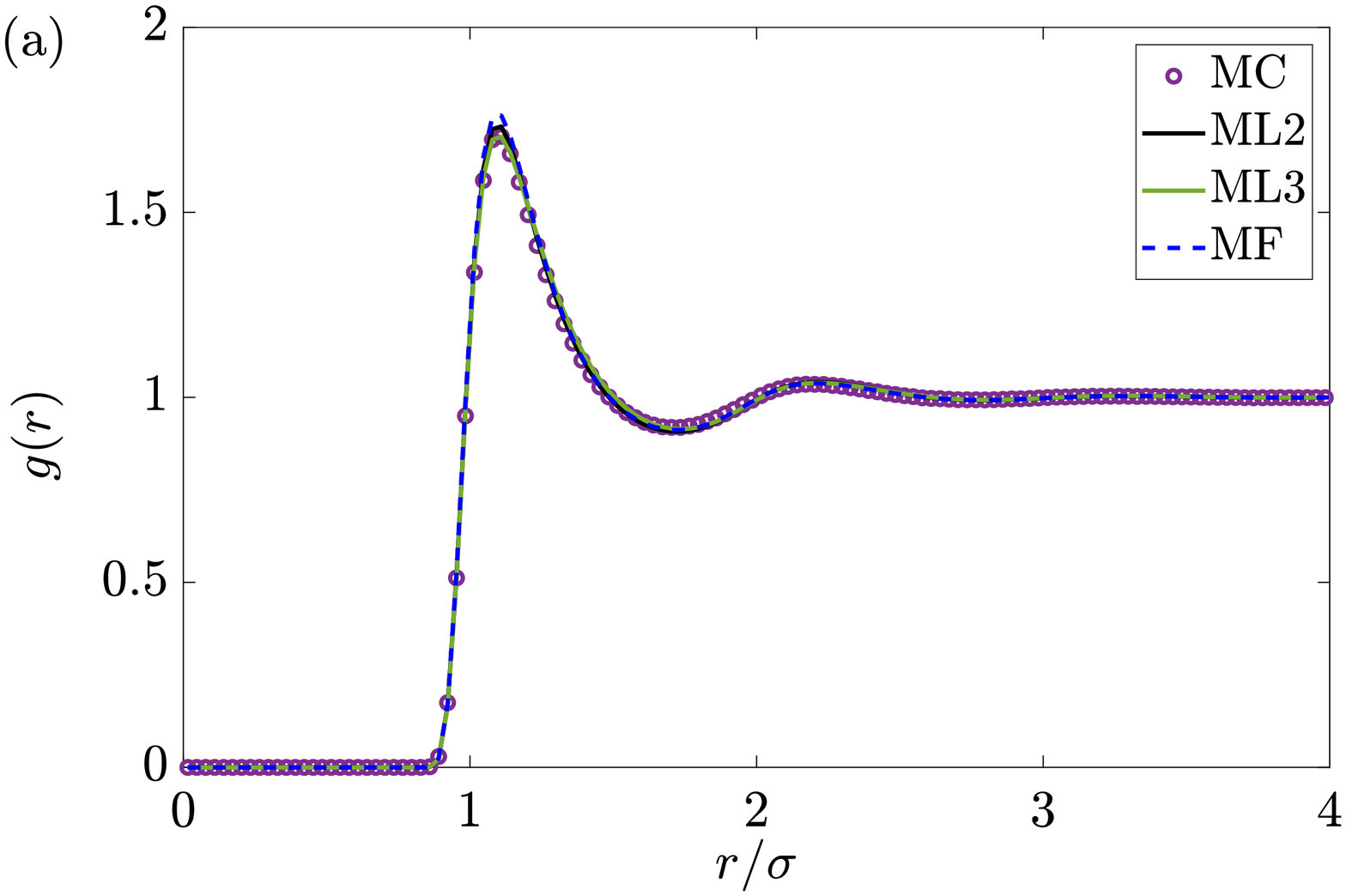}
    \includegraphics[width=0.47\textwidth]{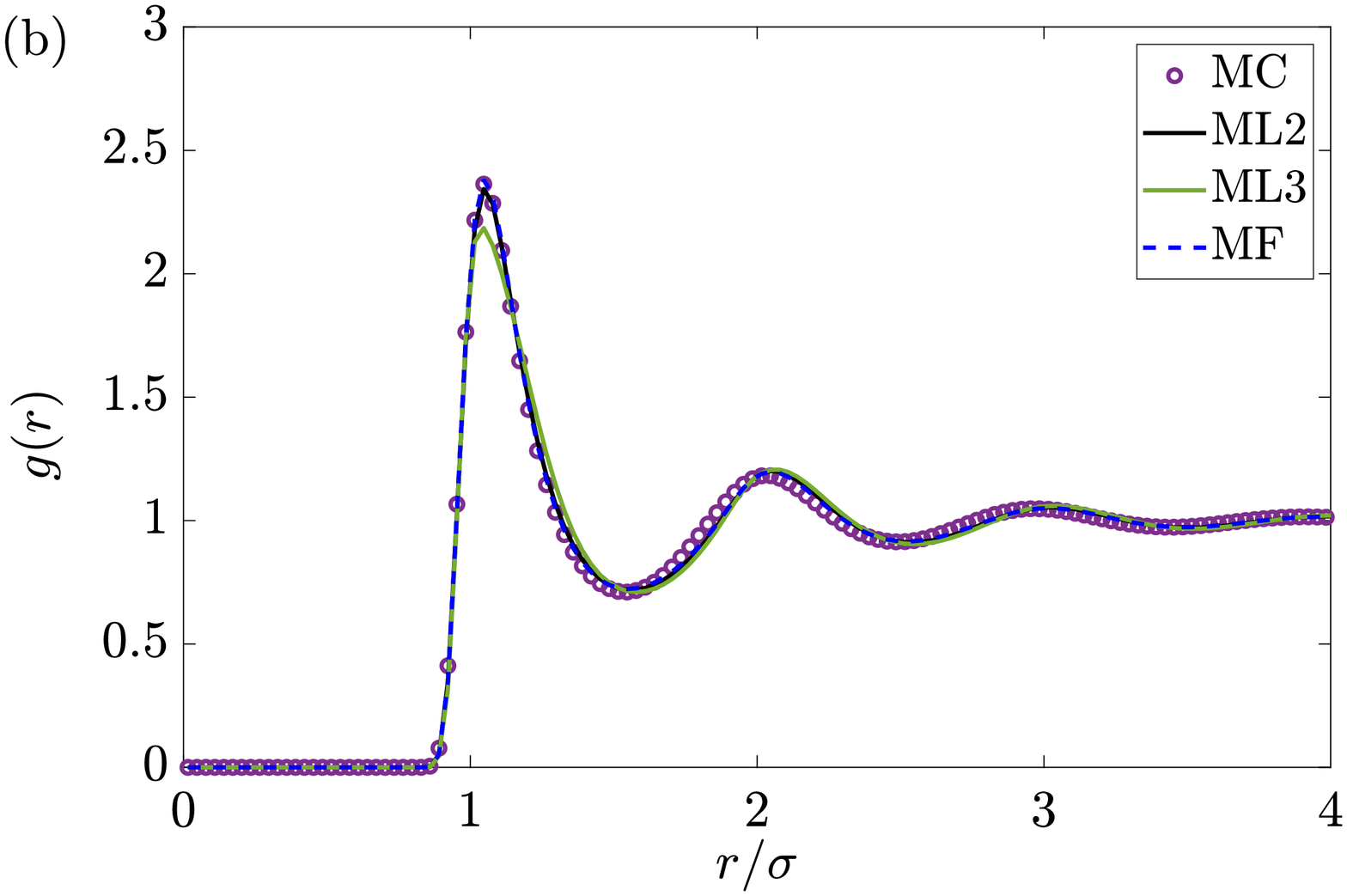}
    \caption{\label{fig:Percustest_lowrho_LJ} The radial distribution function $g(r)$ for the Lennard-Jones system as obtained from the Percus test-particle method for a bulk density (a) $\rho_b\sigma^3 = 0.39$ and (b) $\rho_b\sigma^3 = 0.837$. System, legends, and the MC data are identical to those in Fig.~\ref{fig:OZ_lowrho_LJ}.   }
\end{figure}

Interestingly, DFT provides another procedure to calculate the radial distribution function of a bulk fluid. This so-called ``Percus test-particle method'' \cite{percus1962approximation} is based on the identification of $\rho_bg(r)$ with the equilibrium density profile $\rho_0(r)$ that surrounds a given (test) particle that is fixed in the origin of an otherwise homogeneous fluid at bulk density $\rho_b\equiv\rho_0(\infty)$. In other words, $g(r)=\rho_0(r)/\rho_0(\infty)$ with $\rho_0(r)$ the spherically symmetric density profile of the fluid in an external potential that equals the pair potential, $V_{ext}({\bf r})=\phi(r)$, scaled such that $g(\infty)=1$. For a given chemical potential $\mu$ and a given functional ${\cal F}^{exc}[\rho]$ one thus obtains $g(r)$ through the solution $\rho_0(r)$ of the Euler-Lagrange equation \eqref{eq:selfconsistingequationintroduction}. For the same two state points as used in Fig.~\ref{fig:OZ_lowrho_LJ} we present the resulting radial distributions in Fig.~\ref{fig:Percustest_lowrho_LJ}(a) and (b). For both densities the agreement between simulation and all three DFTs is substantially better than obtained from the Ornstein-Zernike route shown in Fig.~\ref{fig:OZ_lowrho_LJ}, not only for $r<\sigma$ where the Boltzmann factor of the external potential in Eq.~\eqref{eq:selfconsistingequationintroduction} ensures a vanishingly small contribution to $g(r)$ but also at larger distances where the oscillations in the MC data are rather accurately captured by all three DFTs. Interestingly, however, the peaks of the oscillations in (b) are actually better accounted for by MF and ML2 than by ML3, which  underestimates them especially at close contact. The relatively good performance of MF in predicting $g(r)$ via the Percus test-particle method compared to its relatively poor prediction of the equation of state $\rho_b(\mu)$ is due to the scaling-out of $\rho_b$ in the density profile $\rho_0(r)$ such that $g(\infty)=1$ by construction.  Clearly, an overall comparison of Fig.~\ref{fig:OZ_lowrho_LJ} and Fig.~\ref{fig:Percustest_lowrho_LJ} shows that $g(r)$ based on the test-particle method is much more accurate compared to the MC simulations than those based on the Ornstein-Zernike equation.  This is not surprising given that we constructed the direct correlation function in Eq.~\eqref{cb} based on a modification of that of a reference hard-sphere system, which yields a non-vanishing radial distributions inside the hard core if the OZ route is used. A more careful discussion on the different radial distribution functions from the two routes can be found in Ref.~\onlinecite{Archer_2017}.

\section{Summary, discussion, and outlook}
In this article we combine the formalism of classical density functional theory (DFT) with machine learning (ML) density profiles from Monte Carlo (MC) simulations to construct approximations to the excess intrinsic Helmholtz free-energy functional ${\cal F}^{exc}[\rho]$ of a (truncated and shifted)  Lennard-Jones fluid at the supercritical temperature $k_BT/\epsilon=2$. This functional consists of a well-known and accurate hard-sphere contribution, a standard Van der Waals type mean-field account of the attractions, and new  machine-learned corrections  that are, for simplicity, either of a quadratic (ML2) or an additional cubic (ML3) form in the density. The kernels of ML2 and ML3 are radially symmetric and translation-invariant two-point functions of the form $\Omega_i(|{\bf r}-{\bf r}'|)$ for $i=2$ and $3$, see Eqs.~\eqref{eq:ansatz1} and~\eqref{eq:ansatz2}.  By comparing DFT predictions of the equilibrium density profiles $\rho_0(z)$ with grand-canonical MC simulations at a {\sl learning set} of chemical potentials $\mu$ and external potentials $V_{ext}(z)$ in a 3D {\sl planar} geometry, we can construct the optimal planar kernels $\omega_i(|z-z'|)$ using Adam to minimize a suitable loss function, from which we can reconstruct the full {\sl radially} symmetric kernels $\Omega_i(|{\bf r}-{\bf r}'|)$. Given that ${\cal F}^{exc}[\rho]$ is independent from the external potential and the chemical potential, the functional and its Euler-Lagrange equation \eqref{eq:selfconsistingequationintroduction} for $\rho_0({\bf r})$ can be applied to any $\mu$ and any $V_{ext}({\bf r})$. By comparisons with density profiles obtained from grand-canonical MC simulations, for conditions within and outside of the learning set, we find that the ML2 and ML3 functionals generally outperform MF by far because the latter predicts densities that are systematically too low; ML3 improves ML2 somewhat on some of the details at higher $\mu$, at least within the training set. A similar picture emerges from the resulting representations of the mechanical equations of state, {\sl viz.} the bulk density $\rho_b(\mu)$ and the pressure $p(\mu)$, where MF is too low by a large margin and ML3 performs only slightly better than ML2 within the training set, while showing a slightly poorer performance outside. The functional ${\cal F}^{exc}[\rho]$ can also be used to calculate the direct pair correlation function, from which the radial distribution $g(r)$ of a bulk fluid follows via the Ornstein-Zernike equation. At the relatively low bulk density $\rho_b\sigma^3=0.39$ this yields, outside the hard core, an (almost) equally satisfying result for MF, ML2, and ML3, see Fig.~\ref{fig:OZ_lowrho_LJ}. Inside the hard core, and also close to contact, say $\sigma<r<1.3\sigma$, the prediction for $g(r)$ is poor in all cases. However, all three functionals give a rather good account of the simulated $g(r)$ at these two state points if the Percus test-particle method is employed, although here ML3 overestimates the peaks at the higher density slightly. The reason for the relatively good MF performance for $g(r)$ compared to the equation of state $\rho_b(\mu)$ and density profiles stems from the imposed asymptotic normalisation $g(r\rightarrow\infty)=1$. 

A disadvantage of the ML approach is its black-box character, and the associated difficulty to interpret the outcome. In particular the ``hump'' in the ML2 and ML3 kernels $\omega_i(z)$ close to $z=0$ shown in Fig.~\ref{fig:Kernel2lalleTLJ}  and the associated deeply negative direct correlation $c_b(r)$ close to $r=0$ for ML3 in Fig.~\ref{fig:DCF_LJ} are actually rather suspicious. In retrospect, we expect these features to be the result of some degree of overfitting the data in the learning process. This is also borne out by closer inspection of the bulk equations of state $\rho_b(\mu)$ and $p(\mu)$ in Fig.~\ref{fig:Bulkdensity_LJ}(a) and~(b), respectively,  where ML3 hardly improves upon ML2 in the (hatched) regime of the learning set while performing even poorer outside, and likewise for the $g(r)$ of Figs.~\ref{fig:OZ_lowrho_LJ}(a) and~\ref{fig:Percustest_lowrho_LJ}(a) at the density $\rho_b\sigma^3=0.39$ that lies comfortably in the middle of the training set. Of course, ML3 does outperform ML2 somewhat for the density profiles of Fig.~\ref{fig:DPkernel2LJ}. Nevertheless, some more caution could or should have been exercised in the diversity of the training set of external potentials, perhaps with attractive components and discontinuities. We leave studies along these lines for future work. 

Although there is room for improvement and extensions, we have shown here anyway that it is in principle possible to construct a free-energy functional for an atomic fluid by an ML  process that takes data from grand-canonical MC simulations at a variety of chemical and external potentials, from which further predictions {\sl outside} the training set can be made. Interestingly, even data taken in a planar geometry can suffice to construct the full functional, at least for the (relatively simple) functional forms that we considered here which are linear in the kernels $\Omega_i(|{\bf r}-{\bf r}'|)$; nonlinear forms probably require a different treatment. It is important to realise that we fixed the temperature, and although  ${\cal F}^{exc}[\rho]$ is independent of $\mu$ and $V_{ext}({\bf r})$ it is dependent on $T$, so strictly speaking a new functional is to be constructed at every temperature of interest. We leave the $T$-dependence of the functional to future work. Another rather straightforward extension is to use the newly constructed functional to calculate the Gibbs adsorption and the tensions of wall-fluid interfaces, for which we expect good accuracy on the basis of the good agreement of the density profiles.  We also expect that it is possible to extend studies of this type to other systems with spherically symmetric particles, also to mixtures such as electrolytes. Systems of particles with orientation degrees of freedom are probably challenging in practise because of their larger number of variables, although one could imagine first attempts based on truncated expansions in spherical harmonics or an initial focus on homogeneous bulk states (nematics). We hope that this paper will stimulate further explorations of the combination of DFT, ML, and MC simulation.

\section*{acknowledgement} 
This work is part of the D-ITP consortium, a program of the Netherlands Organisation for Scientific Research (NWO) that is funded by the Dutch Ministry of Education, Culture and Science (OCW). It also forms part of the NWO program `Data-driven science for smart and sustainable energy research', with project numbers 16DDS003 and 16DDS014.

Data will be made available upon request.


\nocite{}
\bibliography{biblio}
\end{document}